\documentclass[twocolumn,showpacs,preprintnumbers,amsmath,amssymb]{revtex4}
\usepackage{graphicx}
\usepackage{dcolumn}
\usepackage{bm}

\usepackage{mathptmx}

\newcounter{saveeqn}

\begin{document}

\title{\emph{First-principles} studies of ground- and excited-state properties of MgO, ZnO, and CdO polymorphs}

\author{A. Schleife, F. Fuchs, J. Furthm\"uller, and F. Bechstedt}
\affiliation{Institut f\"ur Festk\"orpertheorie und -optik,
Friedrich-Schiller-Universit\"at, Max-Wien-Platz 1, 07743 Jena,
 Germany }
\date{\today}
\begin{abstract}
An \emph{ab initio} pseudopotential method based on density
functional theory, generalized gradient corrections to exchange
and correlation, and projector-augmented waves is used to
investigate structural, energetical, electronic and optical
properties of MgO, ZnO, and CdO in rocksalt, cesium chloride,
zinc blende, and wurtzite structure. In the case of MgO we also examine
the nickel arsenide structure and a graphitic phase. The stability
of the ground-state phases rocksalt (MgO, CdO) and wurtzite (ZnO)
against hydrostatic pressure and biaxial strain is studied.
We also present the band structures of all polymorphs as well as 
the accompanying 
dielectric functions. We discuss the physical
reasons for the anomalous chemical trend of the ground-state
geometry and the fundamental gap with the size of the group-II
cation in the oxide. The role of the shallow Zn3$d$ and Cd4$d$
electrons is critically examined.
\end{abstract}
 \maketitle

\section{Introduction}

Currently the optical properties of wide band-gap semiconductors such as
zinc oxide (ZnO) are of tremendously increasing
interest, in response to the industrial demand for optoelectronic
devices operating in the deep blue or ultraviolet spectral region.
Under ambient conditions ZnO is crystallizing in wurtzite ($w$)
structure. Its tendency to be grown as fairly high residual
$n$-type material illustrates the difficulty to achieve its
$p$-type doping. Nevertheless, the potential of ZnO for
optoelectronics \cite{1} but also for spintronics \cite{2} (e.g. in
combination with MnO) renders it among the most fascinating 
semiconductors now and of the near future.

Heterostructures with other materials are most important for
optoelectronic applications \cite{B,4}. One important class of
crystals for heterostructures with ZnO could be the other group-II
oxides and alloys with these compounds. Another IIB oxide is CdO
which however has a much smaller fundamental energy gap \cite{5}.
On the other hand, the group-IIA oxide with the smallest cation,
MgO, possesses a much larger energy gap \cite{6}. Consequently,
combinations of these oxides with ZnO should lead to type-I
heterostructures with ZnO (CdO) as the well material and MgO (ZnO)
as the barrier material. However, there are at least two problems
for the preparation of heterostructures: (i) The group-II oxides
MgO, ZnO, and CdO in their ground-states do not represent an
isostructural series of compounds with a common anion, that means the
ground-states are given by different crystal structures: the cubic
rocksalt (MgO, CdO) or the hexagonal wurtzite (ZnO) one \cite{5}. (ii)
The in-plane lattice constants of two oxides grown along the cubic
[111] or hexagonal [0001] direction are not matched.

Among several other properties of the group-II oxides that are not
well understood, is their behavior under hydrostatic and biaxial
strain as well as their stability: Which are possible high-pressure
phases? Are there any strain-induced phase transitions? Which role do
the $d$-electrons play for ZnO and CdO? Other open questions concern
the influence of the atomic geometry on the band structure and,
hence, on the accompanying optical properties. Furthermore, the band-gap
anomaly of CdO has to be clarified: While the band gaps of the
common-cation systems CdTe, CdSe, CdS, and CdO show an increase of
the fundamental gap along the decreasing anion size until CdS, the
value for CdO is smaller than that of CdS \cite{7}. A similar 
anomaly occurs along the row ZnTe, ZnSe, ZnS, and ZnO \cite{7},
however, the gap variation is much smaller. Such gap anomalies are
also observed for III-V semiconductors with a common cation: one
example is the row InSb, InAs, InP, and InN \cite{8}, where the anomaly
for InN has been traced back to the extreme energetical lowering
of the N2$s$ orbital with respect to the Sb5$s$, As4$s$, and P3$s$
levels and the small band-gap deformation potential of InN
\cite{8}.

In this paper, we report well-converged \emph{ab initio}
calculations of the ground- and excited-state properties of the
most important polymorphs of the group-II oxides MgO, ZnO, and
CdO. In Sec. II the computational methods are described.
Atomic geometries and the energetics are presented in Sec. III for
cubic and hexagonal crystal structures.
In Sec. IV we discuss the corresponding band structures
and the electronic dielectric functions. Finally, a brief summary and
conclusions are given in Sec. V.

\section{Computational Methods}

Our calculations are based on the density functional theory (DFT)
\cite{9} in local density approximation (LDA) with generalized gradient
corrections (GGA) \cite{10} according to Perdew and Wang (PW91) \cite{11}.
The electron-ion interaction is described by
pseudopotentials generated within the projector-augmented wave (PAW) scheme
\cite{64,13,14} as implemented in the Vienna Ab initio Simulation Package (VASP)
\cite{12}. The PAW method allows for the accurate treatment of the
first-row element oxygen as well as the Zn3$d$ and Cd4$d$
electrons at relatively small plane wave cutoffs.\\
For the expansion
of the electronic wave functions we use plane waves up to kinetic
energies of 29.4~Ry (MgO in wurtzite, rocksalt, CsCl, NiAs, and
graphitic-like structure, ZnO in zinc blende and wurtzite structure,
and CdO in rocksalt and wurtzite structure) and 33.1~Ry 
(MgO in zinc-blende structure, ZnO in CsCl and 
rocksalt structure, and CdO in CsCl and zinc-blende structure),
respectively. To obtain converged results for the
external pressures (trace of the stress tensor) we increased the plane
wave energy-cutoff to 51.5 Ry uniformly for all materials.
The Brillouin-zone (BZ) integrations in the
electron density and the total energy are replaced by summations
over special points of the Monkhorst-Pack type \cite{15}. We use
8$\times$8$\times$8 meshes for cubic systems and
12$\times$12$\times$7 for hexagonal polymorphs.

A first approach to qualitatively
reliable band structures is using
the eigenvalues of the Kohn-Sham (KS) equation \cite{10}. They
also allow the computation of the electronic density of states
(DOS). We apply the tetrahedron method \cite{16} to perform the
corresponding BZ integration with ${\bf k}$-space meshes
20$\times$20$\times$20 for cubic crystals or
30$\times$30$\times$18 for hexagonal structures.
For the cubic polymorphs the frequency-dependent complex dielectric
function $\varepsilon(\omega)$ is a scalar, but it possesses two
independent tensor components $\varepsilon_{\rm
  xx}(\omega)=\varepsilon_{\rm yy}(\omega)$ and $\varepsilon_{\rm
  zz}(\omega)$ in the cases of hexagonal systems. In independent-particle 
approximation it can be calculated from the Ehrenreich-Cohen formula
\cite{24}. For the BZ integration in this formula we use refined ${\bf k}$-point meshes of 
50$\times$50$\times$31 for hexagonal structures and
40$\times$40$\times$40 for cubic crystals. In
particular, the frequency region below the first absorption peak
in the imaginary part depends sensitively on the number and
distribution of the {\bf k} points. The resulting spectra have been lifetime-broadened
by 0.15~eV but are converged with respect to the used {\bf k}-point meshes.

\section{Ground-state Properties}
\subsection{Equilibrium phases}

For the three oxides under consideration, MgO, ZnO, and CdO we
study three cubic polymorphs: the B1 rocksalt ($rs$ or NaCl)
structure with space group Fm3m $\left(\text{O}^5_h\right)$, the B3 zinc-blende
($zb$ or ZnS) structure with space group F$\bar{4}$3m $\left(\text{T}^2_d\right)$
and the B2 CsCl structure with Pm3m $\left(\text{O}^1_h\right)$. In the case
of the hexagonal crystal system we focus our attention to the
B4 wurtzite ($w$) structure with space group P6$_3$mc $\left(\text{C}^4_{6v}\right)$
but we also investigate the B8$_1$ NiAs
structure with P6$_3$/mmc $\left(\text{D}^4_{6h}\right)$ symmetry and a
graphitic-like structure with the same space group \cite{25} for MgO,
which we call $h$-MgO, according to \cite{62}.
The fourfold-coordinated $w$ and $zb$ structures are
polytypes with the same local tetrahedral bonding geometry, but they differ with
respect to the arrangement of the bonding tetrahedrons in [0001] or [111] direction. Their
high-pressure phases could be the sixfold-coordinated NaCl or
eightfold-coordinated CsCl structures. In the cubic $zb$ and $rs$
phases the cation and anion sublattices are displaced against each
other by different distances parallel to a body diagonal,
$(1,1,1)a_0/4$ for $zb$ and $(1,1,1)a_0/2$ for $rs$, respectively.
There are also several similarities for the three hexagonal phases
wurtzite, NiAs and $h$-MgO. In the NiAs structure the sites of
two ions are not equivalent. For the ideal
ratio $c/a=\sqrt{8/3}$ of the lattice constants the anions (As)
establish a hexagonal close-packed (hcp) structure, whereas the
cations (Ni) form a simple hexagonal (sh) structure. Each cation
has four nearest anion neighbors, whereas each anion has six
nearest neighbors, four cations and two anions. The latter ones
form linear chains parallel to the $c$-axis. In the case of $h$-MgO the
cations and anions form a graphitic-like structure \cite{27} with
B$_k$ BN symmetry. In comparison to wurtzite that leads to flat bilayers 
because of a larger $u$ parameter and an additional layer-parallel mirror plane.
Furthermore the lattice constant $c$ is only somewhat larger than
the in-plane nearest neighbor distance, so that $h$-MgO is essentially five fold coordinated.

\begin{figure}
\resizebox{0.45\textwidth}{!}{\includegraphics*{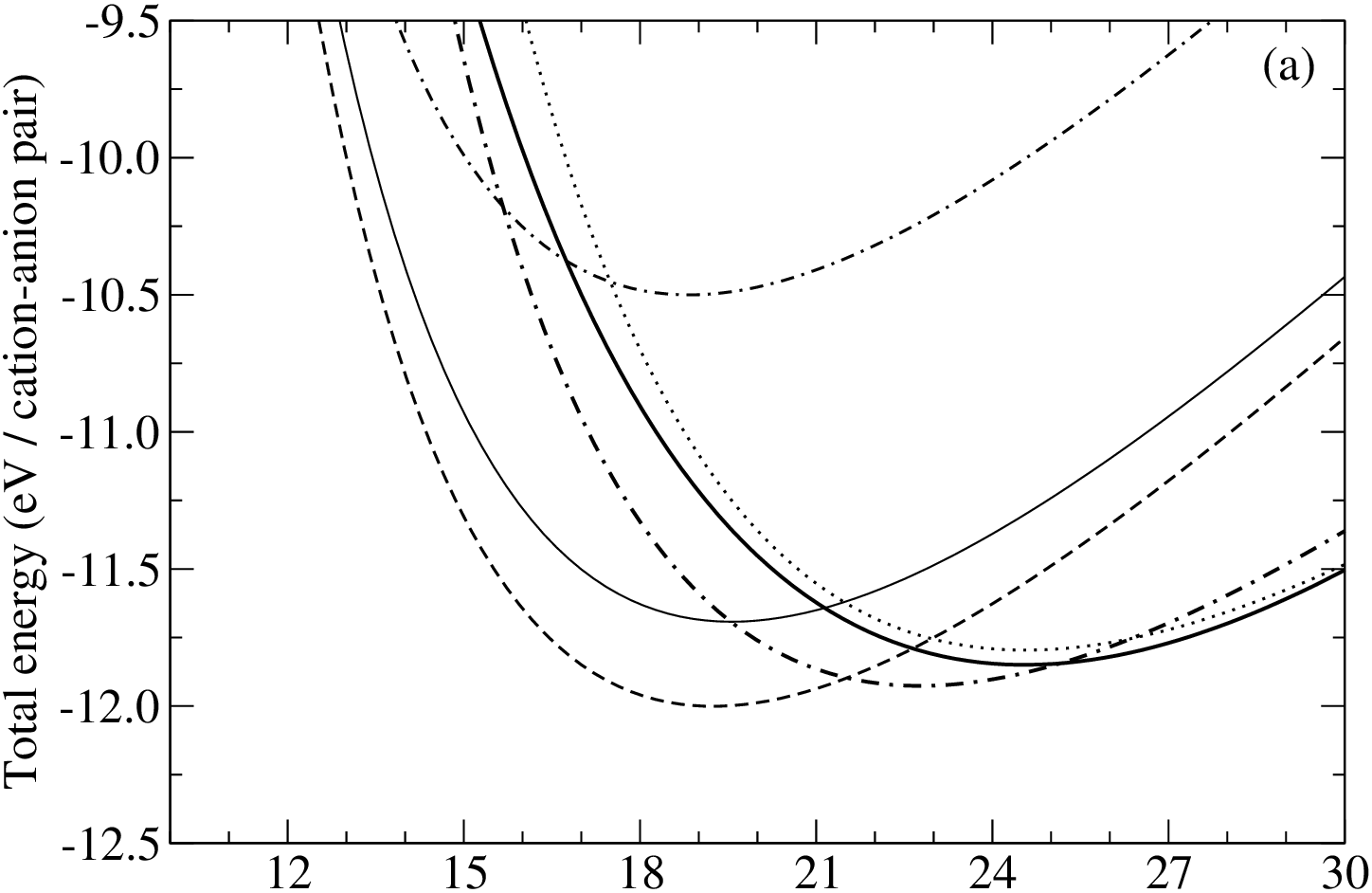}}\\
\resizebox{0.45\textwidth}{!}{\includegraphics*{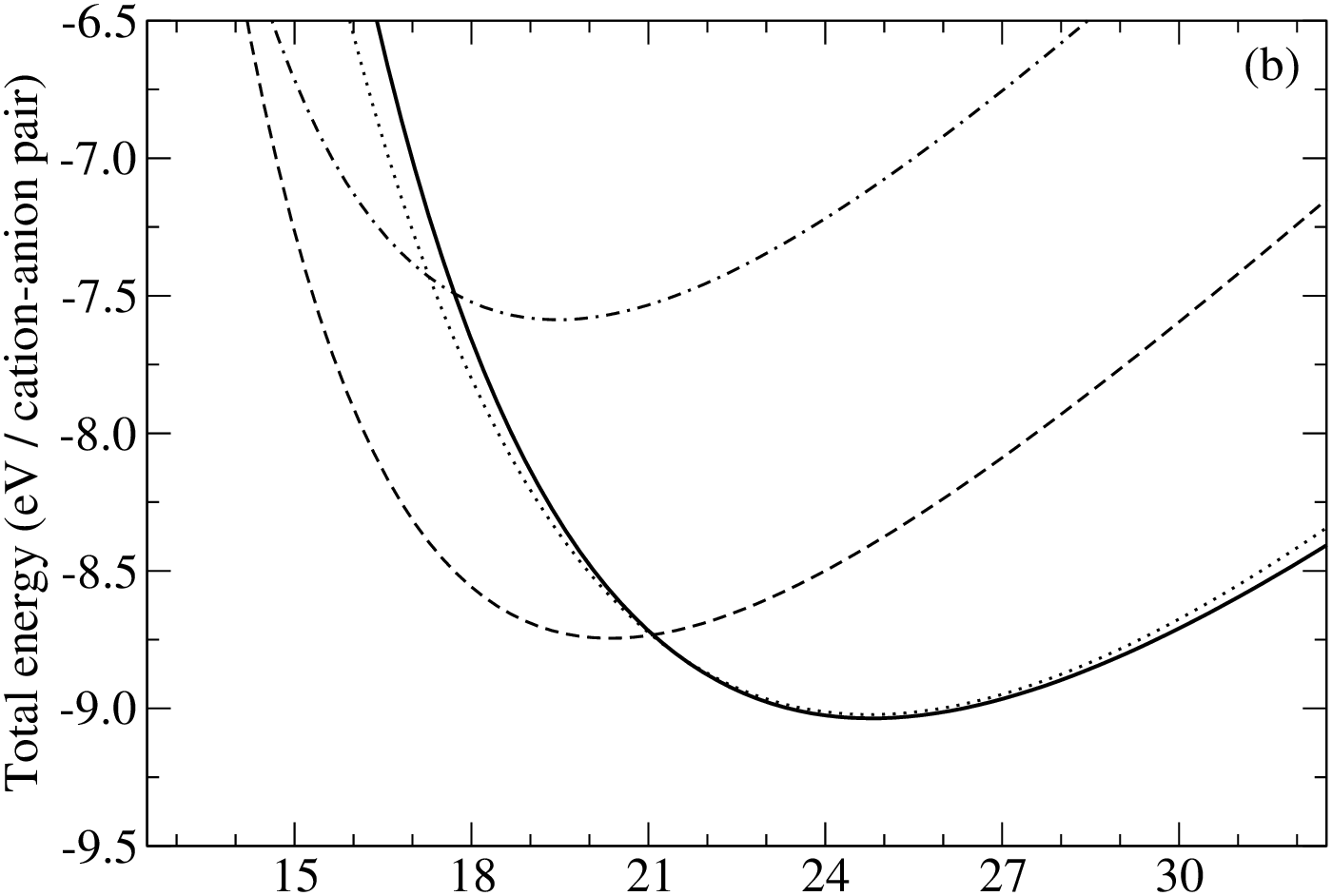}}\\
\resizebox{0.45\textwidth}{!}{\includegraphics*{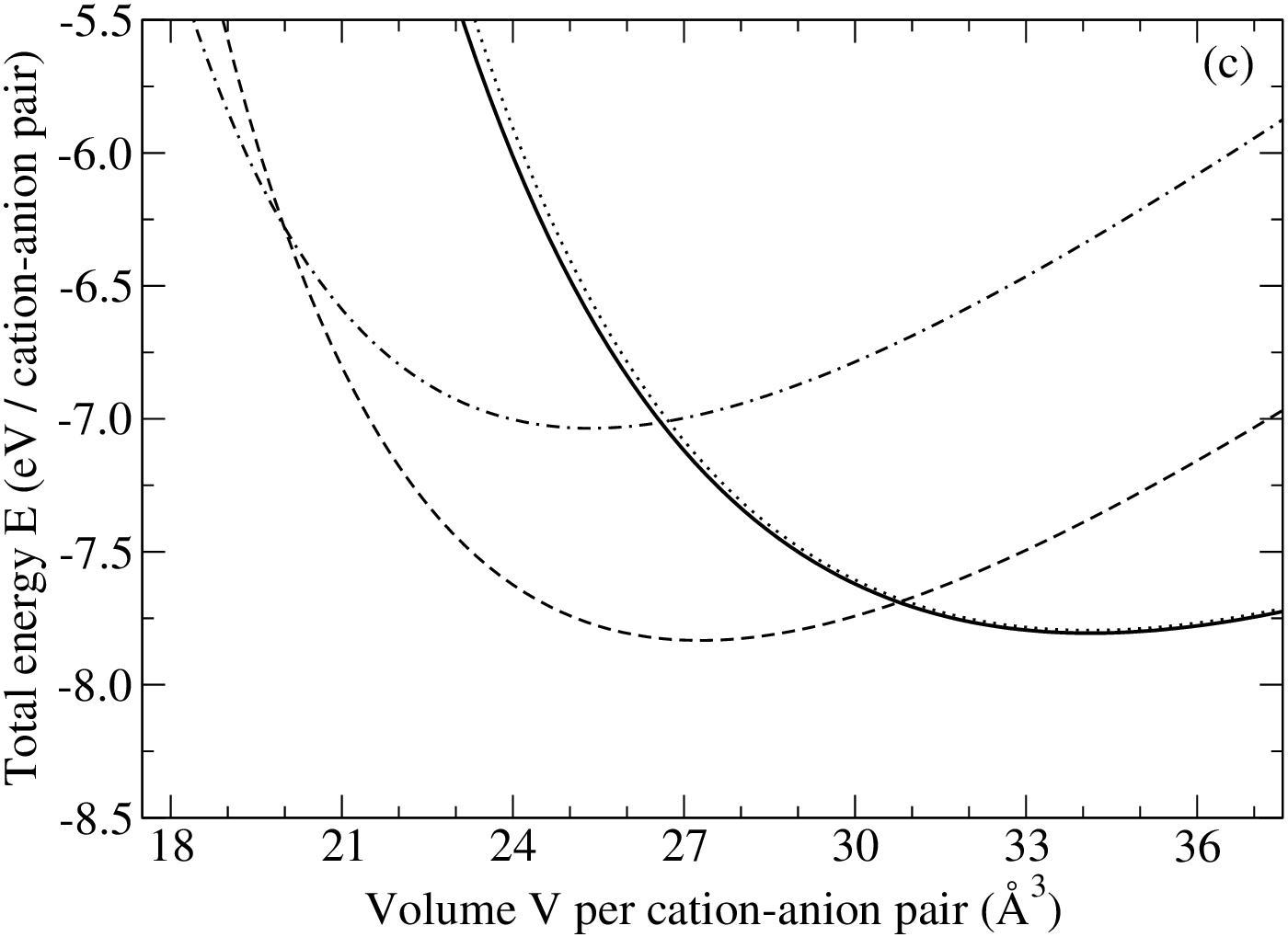}}
\caption{\label{fig:murnag}The nomalized total energy versus volume of one
cation-oxygen pair. Several polymorphs have been studied for MgO
(a), ZnO (b), and CdO (c). $w$: thick solid line, NiAs structure:
thin solid line, $h$-MgO structure: long dash-dotted line, $rs$: dashed
line, $zb$: dotted line, CsCl structure: dash-dotted line }
\end{figure}

To determine the equilibrium lattice parameters the total energy
was calculated for different cell volumes, while the cell-shape
and internal parameters were allowed to relax.
Using the Murnaghan equation of state (EOS) \cite{26} we
obtained the energy-volume dependences $E=E(V)$, the
corresponding fits are represented in Fig.~\ref{fig:murnag}. For the
polymorphs of MgO, ZnO, and CdO being most stable in certain volume
ranges around the equilibrium volumes, these fits lead to the
equilibrium values for the volume $V_0$ and the total energy
$E_0$ per cation-anion pair, as well as the isothermal bulk
modulus $B$, and its pressure coefficient $B'=(dB/dp)_{p=0}$. The
binding energy $E_B$ has been calculated as the corresponding total
energy $E_0$ at zero temperature reduced by atomic total energies computed with
spin polarization. In Table~\ref{tab:ground} these parameters are
summarized together with the lattice parameters we obtained.

\begin{table*}
\caption{\label{tab:ground} Ground-state properties of equilibrium and high-pressure
phases of MgO, ZnO, and CdO. All the experimental binding energies
$^*$) are taken from Ref. \cite{43} as heat of vaporization or
heat of atomization. }
\begin{ruledtabular}
\begin{tabular}{|c|c|ccc|c|c|c|c|}
Oxide & Phase & \multicolumn{3}{c|}{Lattice parameter} & $B$ (GPa) & $B'$ & Binding energy & Reference \\
 & & \multicolumn{3}{c|}{({\AA} or dimensionless)} & & & (eV/pair) & \\ \hline
 MgO & wurtzite & $a=$ & $c=$ & $u=$ & & & & \\
 & & 3.322 & 5.136 & 0.3916 & 116.9 & 2.7 & 10.02 & this work \\
 & & 3.169 & 5.175 & 0.3750 & 137 & -- & 10.85 & theor. \cite{62} \\
 & $h$-MgO & $a=$ & $c=$ & $u=$ & & & & \\
 & & 3.523 & 4.236 & 0.5002 & 124.8 & 4.3 & 10.09 & this work \\
 & & 3.426 & 4.112 & 0.5000 & 148 & -- & 11.09 & theor. \cite{62} \\
 & rocksalt & $a_0=$ & & & & & & \\
 & & 4.254 & & & 148.6 & 4.3 & 10.17 & this work \\
 & & 4.212 & & & -- & -- & \ \ \ 10.26$^*$)  & exp. \cite{land} \\
 & & 4.247 & & & 169.1 & 3.3 & 10.05 & theor. \cite{31} \\
 & & 4.197 & & & 169.0 & 4.2 & -- & theor. \cite{36} \\
 & & 4.253 & & & 150.6 & --  & -- & theor. \cite{42}\\
 & & 4.145 & & & 178.0 & --  & 11.47 & theor. \cite{62}\\
 & cesium chloride & $a_0=$ & & & & & & \\
 & & 2.661 & & & 140.3 & 4.1 & 8.67 & this work \\
 & & 2.656 & & & 152.6 & 3.4 & 8.54 & theor. \cite{31} \\
\hline
 ZnO & wurtzite & $a=$ & $c=$ & $u=$ & & & & \\
 & & 3.283 & 5.309 & 0.3786 & 131.5 & 4.2 & 7.20 & this work \\
 & & 3.258 & 5.220 & 0.382  & 181   & 4   & \ \ \ 7.52$^*$)& exp. \cite{32} \\
 & & 3.250 & 5.204 & --     & 183   & 4   & -- & exp. \cite{28} \\
 & & 3.238 & 5.232 & 0.380  & 154   & 4.3 & -- & theor. \cite{32} \\
 & & 3.292 & 5.292 & 0.3802 & 133.7 & 3.8 & 7.69 & theor. \cite{31} \\
 & & 3.198 & 5.167 & 0.379  & 159.5 & 4.5 & -- & theor. \cite{37} \\
 & & 3.183 & 5.124 & 0.380 & 162 & -- & 10.64 & theor. \cite{44}
 \\
 & zinc blende & $a_0=$ & & & & & & \\
 & & 4.627 & & & 131.6 & 3.3 & 7.19 & this work \\
 & & 4.633 & & & 135.3 & 3.7 & 7.68 & theor. \cite{31} \\
 & & 4.504 & & & 160.8 & 5.7 & -- & theor. \cite{37} \\
 & rocksalt & $a_0=$ & & & & & &  \\
 & & 4.334 & &  & 167.8 & 5.3 & 6.91 & this work \\
 & & 4.275 & &  & 194   & 4.8 & -- & exp. \cite{38} \\
 & & 4.287 & &  & 218   & 4   & -- & exp. \cite{32} \\
 & & 4.271 & &  & 228   & 4   & -- & exp. \cite{28} \\
 & & 4.283 & &  & 202.5 & 3.5 & -- & exp. \cite{30} \\
 & & 4.272 & &  & 198   & 4.6 & -- & theor. \cite{32} \\
 & & 4.345 & &  & 172.7 & 3.7 & 7.46 & theor. \cite{31} \\
 & & 4.316 & &  & 175   & 5.4 & -- & theor. \cite{38} \\
 & & 4.225 & &  & 209.1 & 2.7 & -- & theor. \cite{37} \\
 & & 4.213 & &  & 210   & -- & 10.43 & theor. \cite{44} \\
 & cesium chloride & $a_0=$ & & & & & & \\
 & & 2.690 & & & 162.4 & 4.7 & 5.76 & this work \\
 & & 2.705 & & & 156.9 & 3.8 & 6.33 & theor. \cite{31} \\ \hline
 CdO & wurtzite & $a=$ & $c=$ & $u=$ & & & & \\
 & & 3.678 & 5.825 & 0.3849 & 92.7 & 4.7 & 5.97 & this work \\
 & & 3.660 & 5.856 & 0.3500 & 86 & 4.5 & 5.30 & theor. \cite{39} \\
 & zinc blende & $a_0=$ & & & & & & \\
 & & 5.148 & & & 93.9 & 5.0 & 5.96 & this work \\
 & & 5.150 & & & 82   & 3.0 & 5.18 & theor. \cite{39} \\
 & rocksalt & $a_0=$ & & & & & & \\
 & & 4.779 & & & 130.5 & 5.0 & 6.00 & this work \\
 & & 4.696 & & & 148 & 4 & \ \ \ 6.40$^*$) & exp. \cite{40} \\
 & & 4.770 & & & 130 & 4.1 & 5.30 & theor. \cite{39}\\
 \end{tabular}
\end{ruledtabular}
\end{table*}

The $E(V)$ curves in Fig.~\ref{fig:murnag} clearly show that under ambient
conditions the group-II oxides crystallize either in $rs$ (MgO
and CdO) or $w$ (ZnO) structure. However, the binding energies of
the atoms in $rs$ and $w$ structure are rather similar, especially for CdO. 
The energy gains due to electrostatic attraction 
on smaller distances (NaCl) and due to the better overlap of
$sp^3$ hybrids (wurtzite) result in a sensitive energy
balance.
Therefore we cannot give a simple explanation,
why one of these crystal structures has to be favored over the other
one.
Furthermore we observe an anomalous structural trend along the cation
row Mg, Zn, and Cd, which follows the anomalous trend of their
covalent radii 1.36, 1.25, and 1.48~{\AA} \cite{A}. Together with the
oxygen radius of 0.73~{\AA} the resulting nearest-neighbor distances
2.09, 1.98, and 2.21~{\AA} in the tetrahedrally coordinated wurtzite structure
take a minimum for ZnO. That means besides the strong covalent
bonds due to $sp^3$ hybrid overlapping, also significant energy
gain due to the Madelung energy occurs. As a result of the \emph{ab
  initio} calculations (cf. Table~\ref{tab:ground}) the
nearest-neighbor distances in the sixfold-coordinated rocksalt
structure show a minor monotonous variation 2.13, 2.17, and
2.39~{\AA}. Also the corresponding ionic energies, the repulsive
interaction and the Madelung energy follow a monotonous trend. The
sequence of the ratios of the nearest-neighbor distances of these two
polymorphs with about 0.98, 0.91, and 0.93 may also be considered as
an indication why ZnO exhibits another equilibrium structure as MgO
and CdO and, hence, for the nonexistence of an isostructural
series. Both, the favorization of $w$ or $rs$ structure, as well as
the cationic trend we observed, may be explained with the ideas of
Zunger \cite{63}. Within his model he examined over 500 different
compounds and predicted the correct equilibrium structures for the
three materials we investigated.

Comparing the results with data of recent measurements
or other first-principles calculations, we find
excellent agreement (cf. Table~\ref{tab:ground}). 
For ZnO the wurtzite ground-state and the NaCl (and CsCl)
high-pressure phases are confirmed by experimental studies
\cite{28,29,30,32,33} and other \emph{ab initio} calculations
\cite{28,29,31,32,33}. Experimental results for the rocksalt
ground-state of MgO \cite{land} and CdO \cite{42} exist,
as well as other calculations for the equilibrium
structure \cite{31,36,42,39} and the high-pressure phase (CsCl)
\cite{31,39}. In the case of CdO also the CsCl structure has
been studied experimentally \cite{41}.\\
Our computed lattice parameters for the CdO polymorphs show excellent agreement,
in particular compared with the results of Jaffe et al.
\cite{31}. These authors also use a DFT-GGA scheme but expand the
wave functions in localized orbitals of Gaussian form. However,
also the agreement with values from other computations is
excellent. Most of them use a DFT-LDA scheme which tends to an
overbinding effect, i.e., too small lattice constants, too large
bulk moduli, and too large binding energies. The overestimation
of the binding energies of ZnO polymorphs in Ref.
\cite{44} is probably a consequence that not spin-polarized atomic
energies have been substracted.\\
The DFT-GGA scheme we used tends to underestimate slightly the
bonding in the considered group-II oxide polymorphs. This
underestimation results in an overestimation of the lattice
constants of about 1\%. In the case of the $c$-lattice constants
of $w$-ZnO and the $a_0$ constant of $rs$-CdO this discrepancy
increases to roughly 2\%. Also our calculated bulk moduli are always
smaller than the experimental ones, which may be due to the mentioned
underestimation. Besides the limitation of the computations we cannot
exclude that sample-quality problems play a role for these
discrepancies. For $rs$-MgO, $w$-ZnO, and $rs$-CdO the computed
binding energies are close to the measured ones. The theoretical
underestimation only amounts to 1, 5 or 7\% which are small
deviations.

\subsection{Pressure- and strain-induced phase transitions}

\begin{figure}
\resizebox{0.45\textwidth}{!}{\includegraphics*{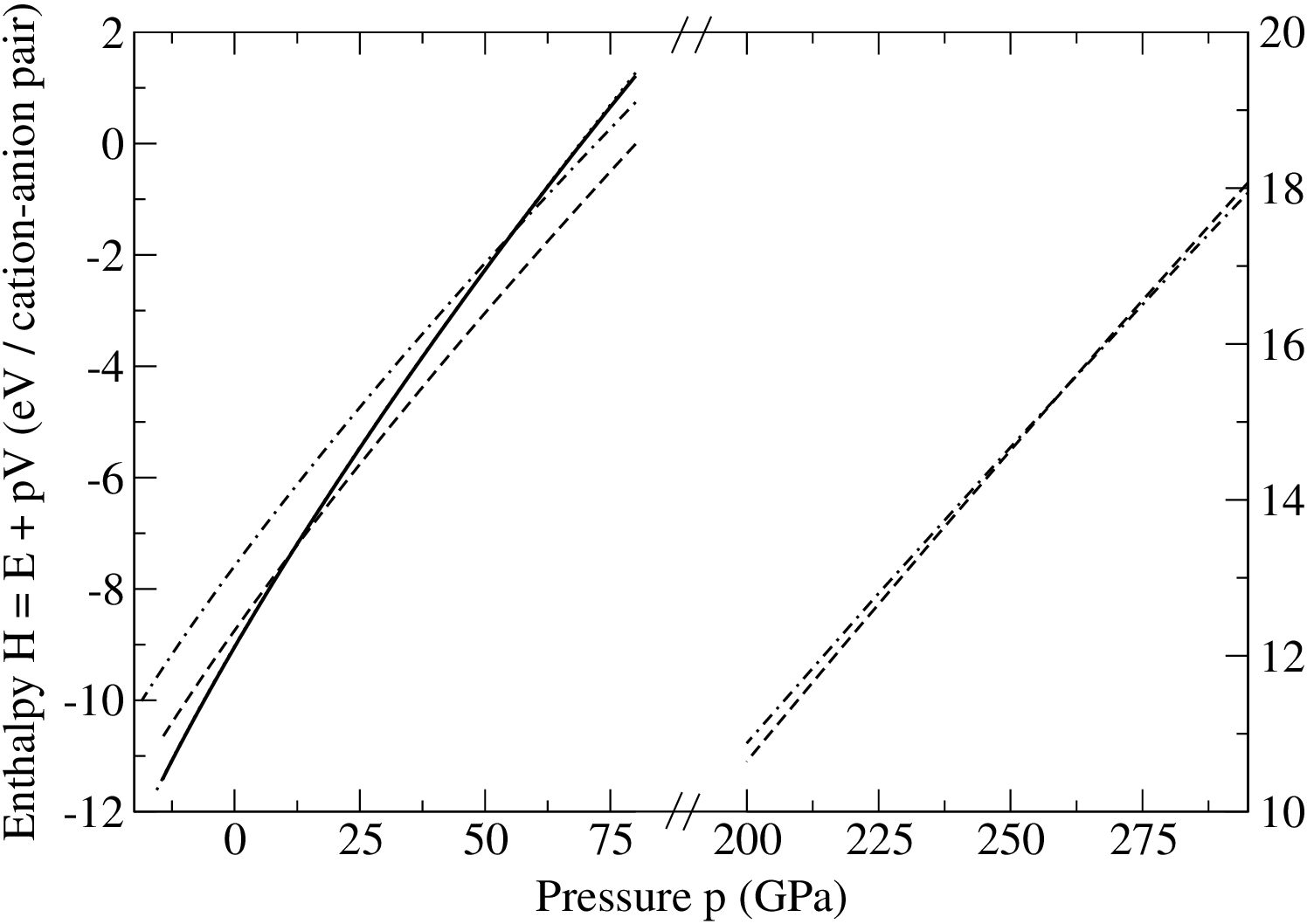}}
\caption{\label{fig:enthalpy} Enthalpy per cation-anion pair of ZnO phases as a
function of hydrostatic pressure. $w$: thick solid line, $rs$:
dashed line, CsCl structure: dash-dotted line. The curve for
zinc blende is practically identical with that of wurtzite.}
\end{figure}

Structural changes in the form of pressure-induced phase
transitions are studied in detail in Fig.~\ref{fig:enthalpy}
for ZnO. Usually the Gibbs free energy $G=U+pV-TS$ as the appropriate thermodynamic
potential governs the crystal stability for given pressure and
temperature. Its study however requires the knowledge of the full
phonon spectrum. Therefore, we restrict ourselves to the
discussion of the low-temperature limit, more strictly speaking to
the electronic contribution to the enthalpy $H=E+pV$ with the internal energy $U(V)\approx E(V)$
and the external pressure obtained as the trace of the stress tensor.
The zero-point motional energy is neglected. Such an approach
is sufficient for the discussion of the pressure-induced
properties of relatively hard materials for temperatures below
that given by the maximum frequency of the phonon spectrum
\cite{34}. For a given pressure the crystallographic phase with the lowest enthalpy
is the most stable one, and a crossing of two
curves indicates a pressure-induced first-order phase transition.\\
From Fig.~\ref{fig:enthalpy} we
derived the equilibrium transition pressure $p_t$. Using $p_t$ we
obtained from $p$ over $V$ plots the initial volume $V_i$ and final volume $V_f$
for the transitions, given here in units of the equilibrium volume $V_0$ of
the wurtzite polymorph. We derive the values $p_t=11.8$~GPa
$(V_i=0.92V_0$, $V_f=0.77V_0)$ for the transition wurtzite
$\rightarrow$ NaCl and 
$p_t=261$~GPa $(V_i=0.50V_0$, $V_f=0.47V_0)$ for the transition NaCl $\rightarrow$
CsCl. The first values are in rough agreement with the experimental findings
$p_t=9.1$~GPa $(V_t=0.82V_0)$ \cite{30}, $p_t\approx 10$~GPa
\cite{29}, or $p_t\approx 9$~GPa \cite{35}, though our calculations
indicate a slightly higher stability of the wurtzite structure
over the rocksalt one. The computed $p_t$ value is in reasonable
agreement with other calculations (see \cite{31} and references
therein). Another pressure-induced phase transition between
NaCl and CsCl structure is found at a transition pressure of
261~GPa, very similar to a previous calculation \cite{31} which
predicted a value of $p_t=256$~GPa.

Applying the common-tangent method the $E(V)$ curves in Figs.~\ref{fig:murnag}a
and~\ref{fig:murnag}c for MgO and CdO already show that pressure-induced phase
transitions are hardly observable or should occur at high
transition pressures (and hence small transition volumes). In the case of MgO the
crossing of the enthalpies gives a value of $p_t=508$~GPa for the
transition NaCl~$\rightarrow$~CsCl structure. For CdO we obtain
$p_t\approx 85$~GPa for the transition NaCl~$\rightarrow$~CsCl
structure. In comparison with the result of a previous calculation \cite{42}
we find good agreement. Another predicted value
amounts to 515~GPa (see \cite{31} and references therein). Also the CdO
value is in excellent agreement with other DFT-GGA calculations $p_t=89$~GPa
\cite{40} and even with the result of measurements $p_t=90$~GPa \cite{41}.\\
For MgO (cf. Fig.~\ref{fig:murnag}a) we have to mention an interesting
result for negative pressures: at large volumes of about
1.3$V_0$, indeed wurtzite is more stable than the NaCl structure.
However, when we start the atomic relaxation with the wurtzite
structure and slightly decrease the cell volume, we observe a
transition into the $h$-MgO structure. The corresponding total energy
minimum therefore lies between the rocksalt and the wurtzite minima at
a volume of about $V=1.2V_0$. Around this volume there is no energy
barrier between the wurtzite and $h$-MgO structures and the wurtzite
geometry only represents a saddle point on the total energy surface,
whereas we observe $h$-MgO to be an intermediate structure on the way
from wurtzite to rocksalt, as discussed in \cite{62}.
There is another indication for this transition: A decrease of $c/a$ leads
to an increase of $u$, followed by a sudden relaxation into the $h$-MgO
structure \cite{71}. Comparing the $u$
parameters of our wurtzite structures (cf. Table~\ref{tab:ground}), we
observe that $w$-MgO has the highest $u$ and so this relaxation is
most probable for MgO.

\begin{figure}
\resizebox{0.45\textwidth}{!}{\includegraphics*{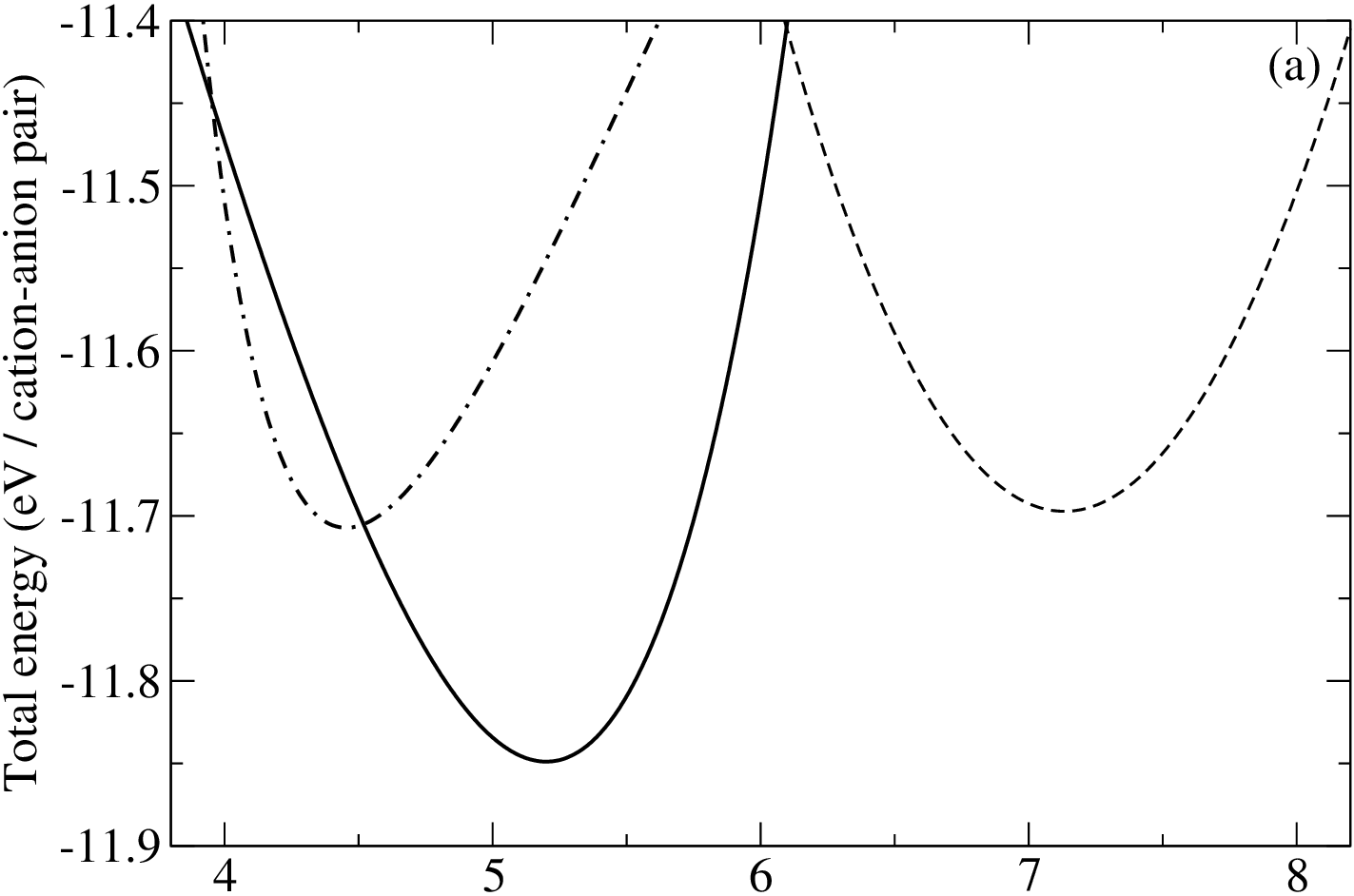}}\\
\resizebox{0.45\textwidth}{!}{\includegraphics*{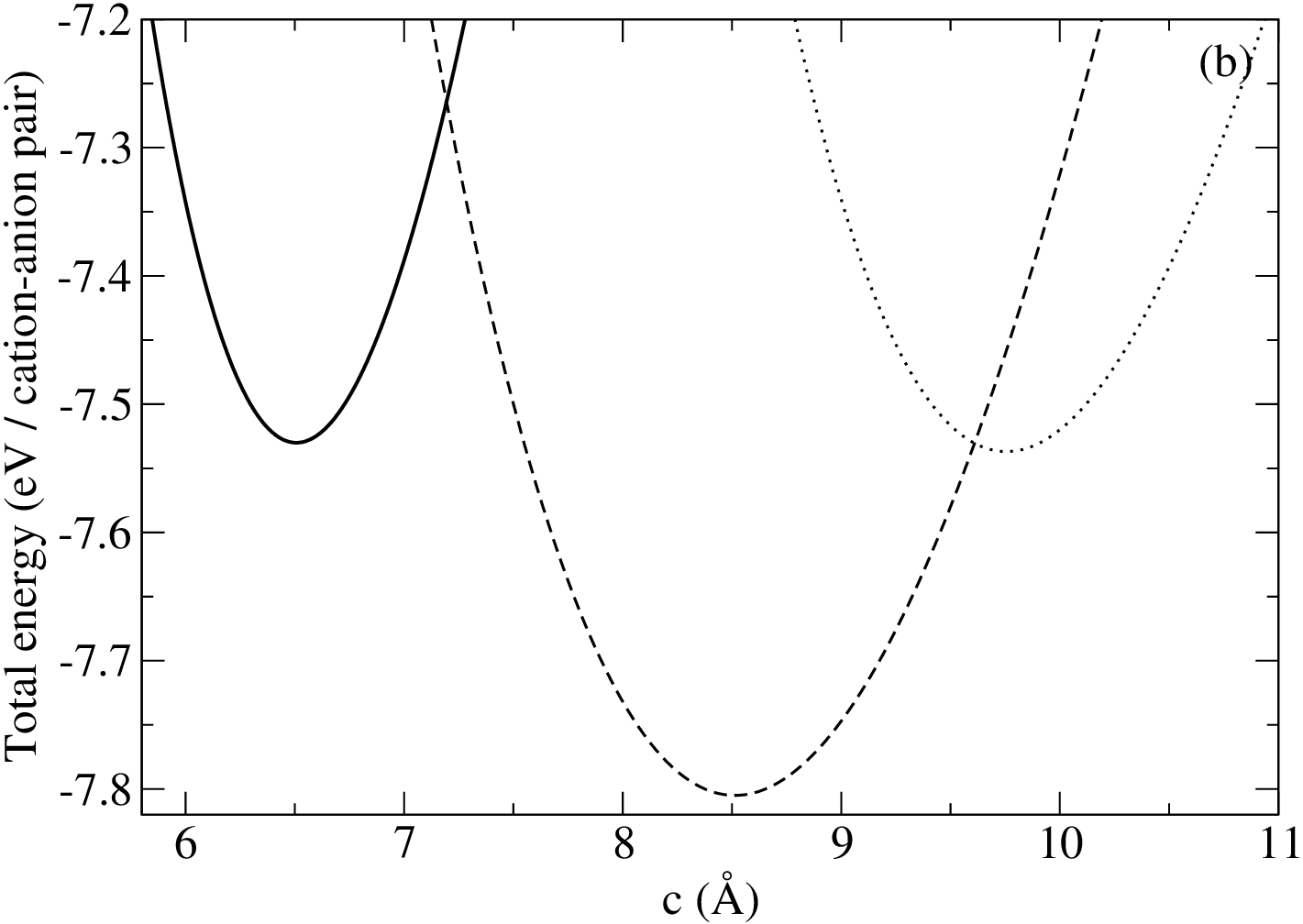}}
\caption{\label{fig:constrained} Total energy of biaxially strained MgO (a) and CdO (b)
polymorphs versus the $c$-lattice constant in [111] or [0001]
direction for an $a$-lattice constant fixed at the value
$a=3.283$~{\AA} of wurtzite ZnO. $w$: thick solid line, $rs$: dashed
line, $zb$: dotted line, and $h$-MgO structure: long dash-dotted line.}
\end{figure}

An important point for the above-mentioned heterostructures it is the
possibility to grow pseudomorphically one material on the
other. Because wurtzite ZnO substrates are commercially available, the
question arises if such growth of certain polymorphs of MgO
or CdO on a $w$-ZnO substrate with [0001] orientation is possible.
For that reason we compare the $a$-lattice constant of $w$-ZnO,
$a=3.283$~{\AA}, with the corresponding lattice constants $a$ of
hexagonal modifications of MgO or CdO and the second-nearest
neighbor distances $a_0/\sqrt{2}$ in the cubic cases. The
corresponding values are 3.523 ($h$-MgO), 3.322 ($w$), and 3.008~{\AA}
($rs$) for MgO or 3.678 ($w$), 3.640 ($zb$), and 3.379~{\AA} ($rs$)
for CdO. With these values the resulting lattice misfits are
7.3, 1.2, and $-8.4$\% for MgO and 12.0, 10.9, and 2.9\% for CdO
for [0001]/[111] interfaces. From the point of
lattice-constant matching pseudomorphic growth of $w$-MgO and,
perhaps, also $rs$-CdO should be possible on a $w$-ZnO substrate.
This conclusion is confirmed by the total-energy studies (see
Fig.~\ref{fig:constrained}) for several MgO and CdO polymorphs
biaxially strained in [0001] or [111] direction of cation-anion bilayer stacking.
For the curves in Fig.~\ref{fig:constrained} we kept the
corresponding $a$-lattice constant fixed at the $w$-ZnO value and computed
the total energy per cation-anion pair for several values of the $c$-lattice
constant of the corresponding hexagonal crystal or of the resulting
rhombohedral crystal, the resulting lattice parameters are given in
Table~\ref{tab:strain}.

\begin{table}
\caption{\label{tab:strain} Lattice parameters of the energetically
  preferred polymorphs of MgO and CdO under biaxial strain along the bilayer-stacking axis.}
\begin{ruledtabular}
\begin{tabular}{|c|c|c|c|c|}
\multicolumn{3}{|c|}{$w$-MgO} & \multicolumn{2}{c|}{$rs$-CdO}\\
\hline
 $a$ (\AA) & $c$ (\AA) & $u$        & $a$ (\AA) & $c$ (\AA) \\ \hline
 3.283     & 5.201     & 0.3856     & 4.643     & 8.512   \\
\end{tabular}
\end{ruledtabular}
\end{table}

The results for MgO are most interesting: In the presence of a
weak biaxial strain in [0001] direction of about 1.3\% the most
energetically favorable geometry is the wurtzite structure with a
resulting $c$-lattice constant of $c=5.21$~{\AA}. The
two polymporphs also considered here, the $h$-MgO structure and the trigonally distorted
$rs$ geometry, are much higher in energy. Furthermore the energetical ordering
of the MgO polymorphs (cf. Fig.~\ref{fig:murnag}a) is completely
changed and MgO adopts the wurtzite crystal
structure of the substrate which is different from its rocksalt
geometry in thermal equilibrium. From the $u$ parameters for
constrained $w$-MgO (see Table~\ref{tab:strain}) and $w$-MgO in
equilibrium (see Table~\ref{tab:ground}) we find that according
to~\cite{71} the biaxially strained wurtzite structure should be more
stable against a transition into $h$-MgO, because its $u$
parameter is closer to its ideal value. We conclude that pseudomorphic
growth of MgO should be possible on ZnO(0001) substrates. We find a somewhat different situation
for CdO. Apart from the large lattice misfit which probably prohibits
pseudomorphic growth, its lowest energy structure is still derived
from the $rs$ atomic arrangement. Of course, there is a strong
trigonal distortion giving rise to a $c$-lattice constant of
$c=8.512$~{\AA} in comparison with the thickness $\sqrt{3}a_0$ of three CdO bilayers in [111] direction
at thermal equilibrium with $c=8.277$~{\AA}. In the case of CdO the two other
crystallographic structures, $w$ and $zb$, are energetically less
favorable also in the biaxially strained case.

\section{Excited-state Properties}
\subsection{Band structures and electronic densities of states}

\begin{figure}
\resizebox{0.39\textwidth}{0.29\textwidth}{\includegraphics*{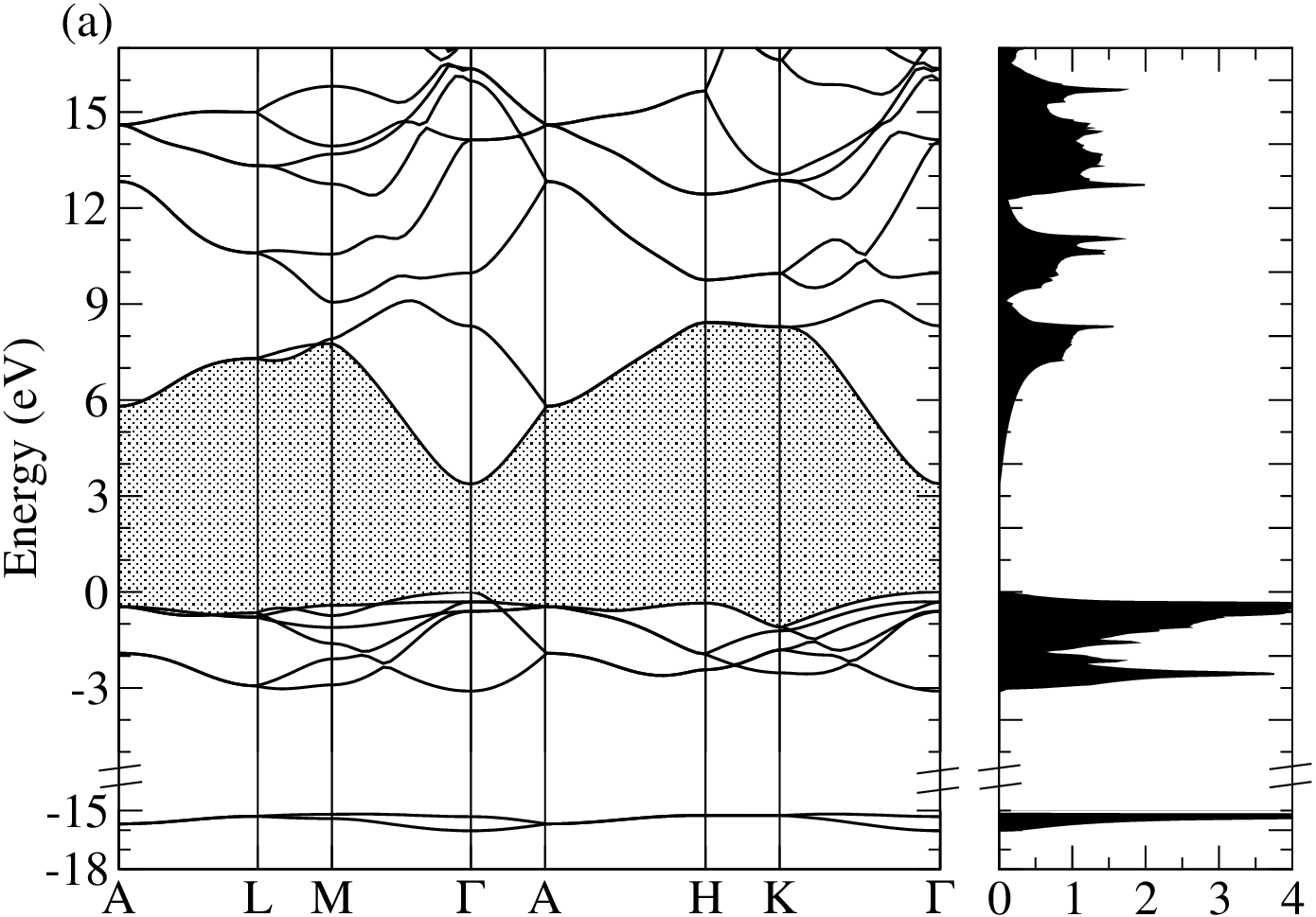}}\\
\vspace{0.1cm}
\resizebox{0.39\textwidth}{0.29\textwidth}{\includegraphics*{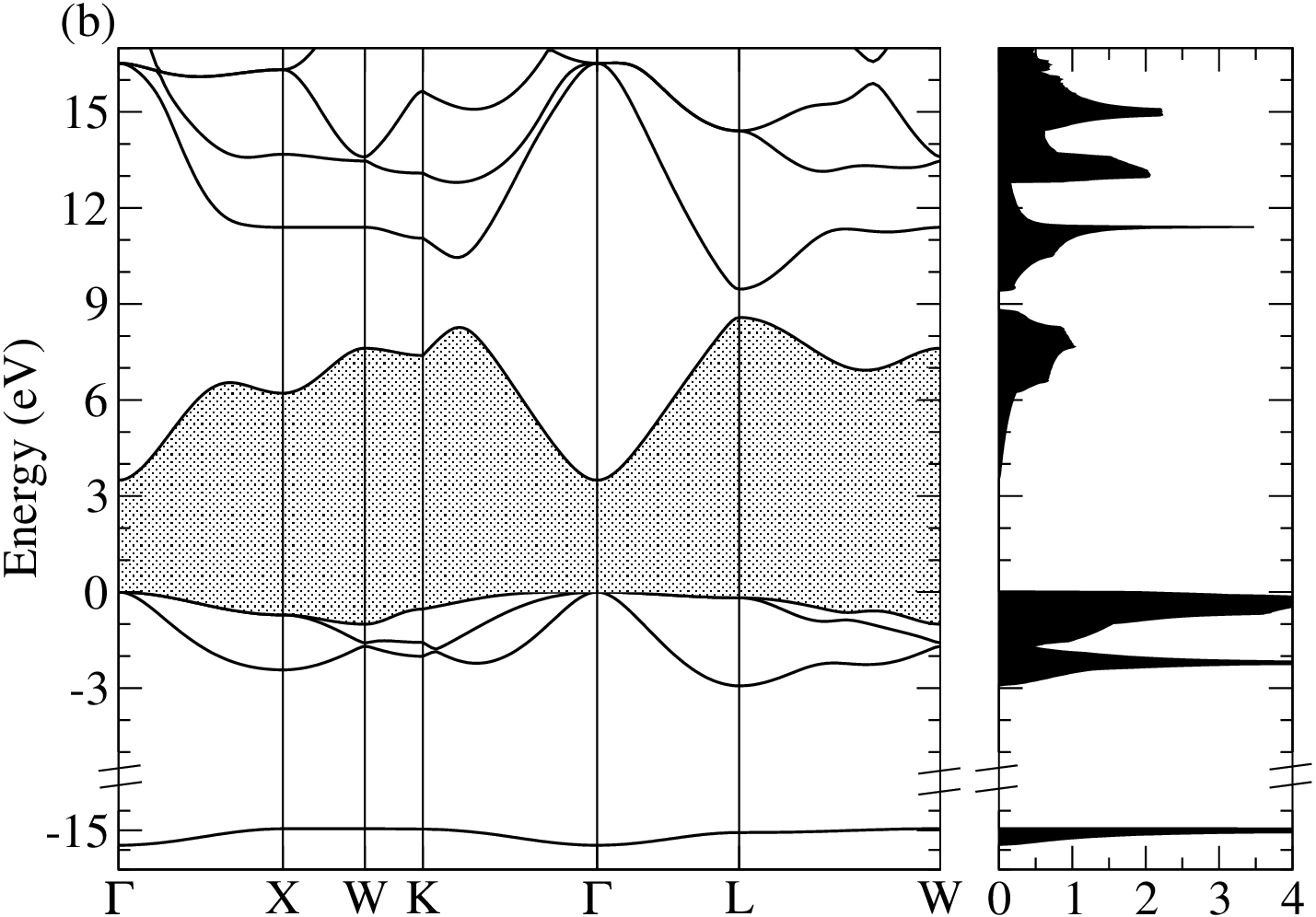}}\\
\vspace{0.1cm}
\resizebox{0.39\textwidth}{0.29\textwidth}{\includegraphics*{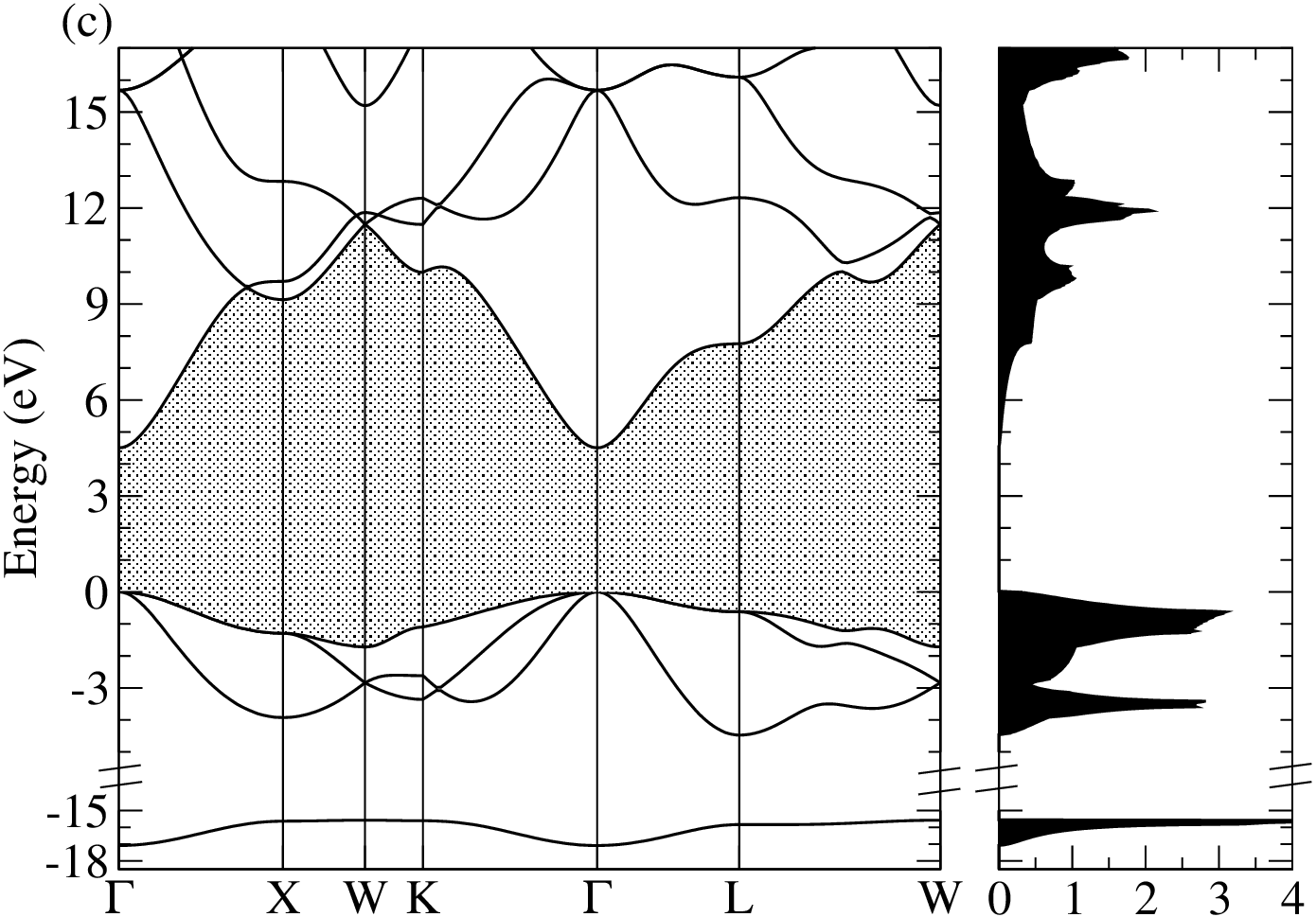}}\\
\vspace{0.1cm}
\resizebox{0.39\textwidth}{0.30\textwidth}{\includegraphics*{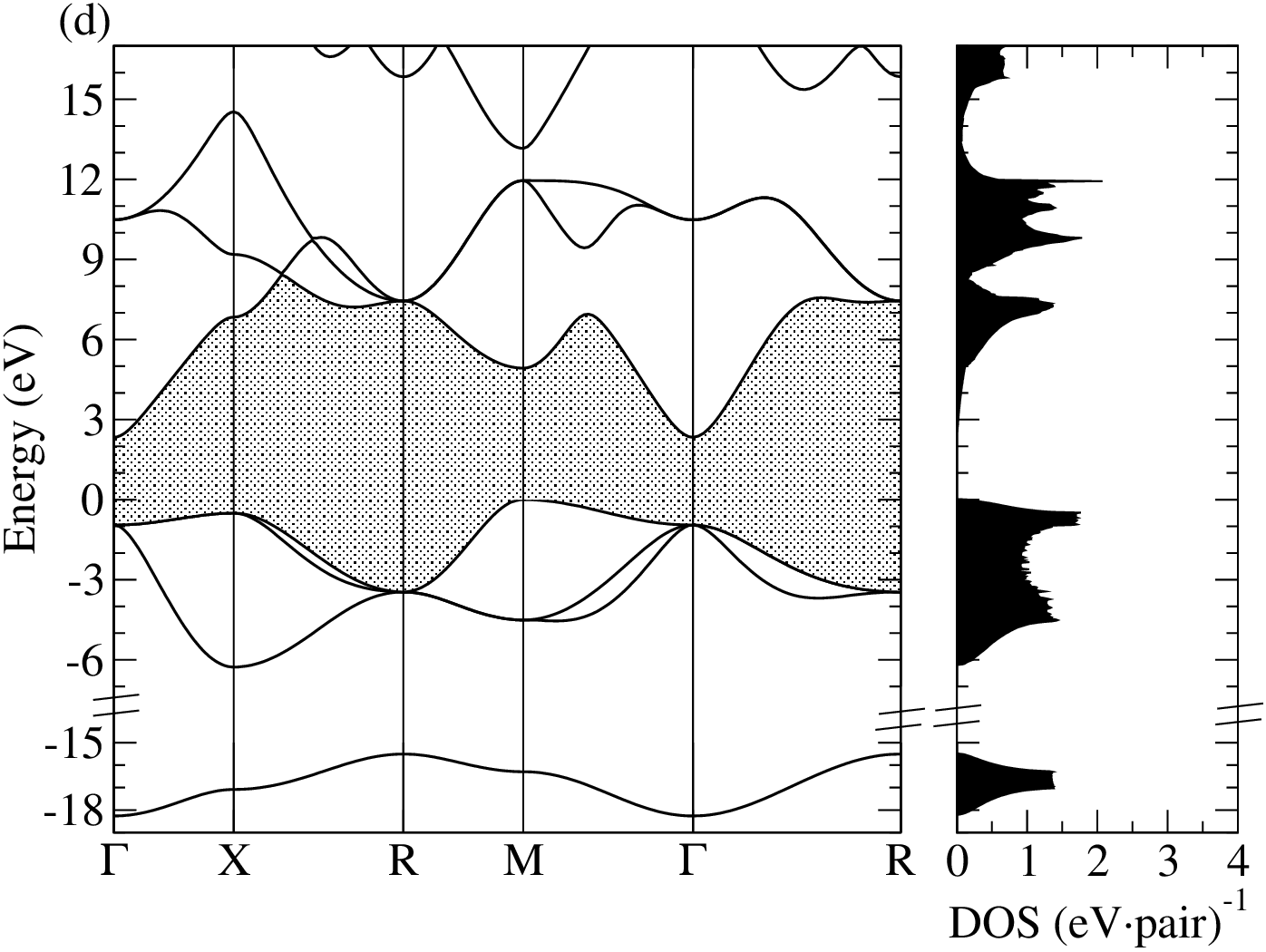}}
\caption{\label{fig:electronic_mgo} Band structure and density of
  states (normalized per pair) for MgO polymorphs 
calculated within the DFT-GGA framework: (a) wurtzite, (b)
zinc blende, (c) rocksalt, and (d) cesium chloride. The shaded region
indicates the fundamental gap. The valence band maximum is chosen as energy zero.}
\end{figure}

\begin{figure}
\resizebox{0.39\textwidth}{0.29\textwidth}{\includegraphics*{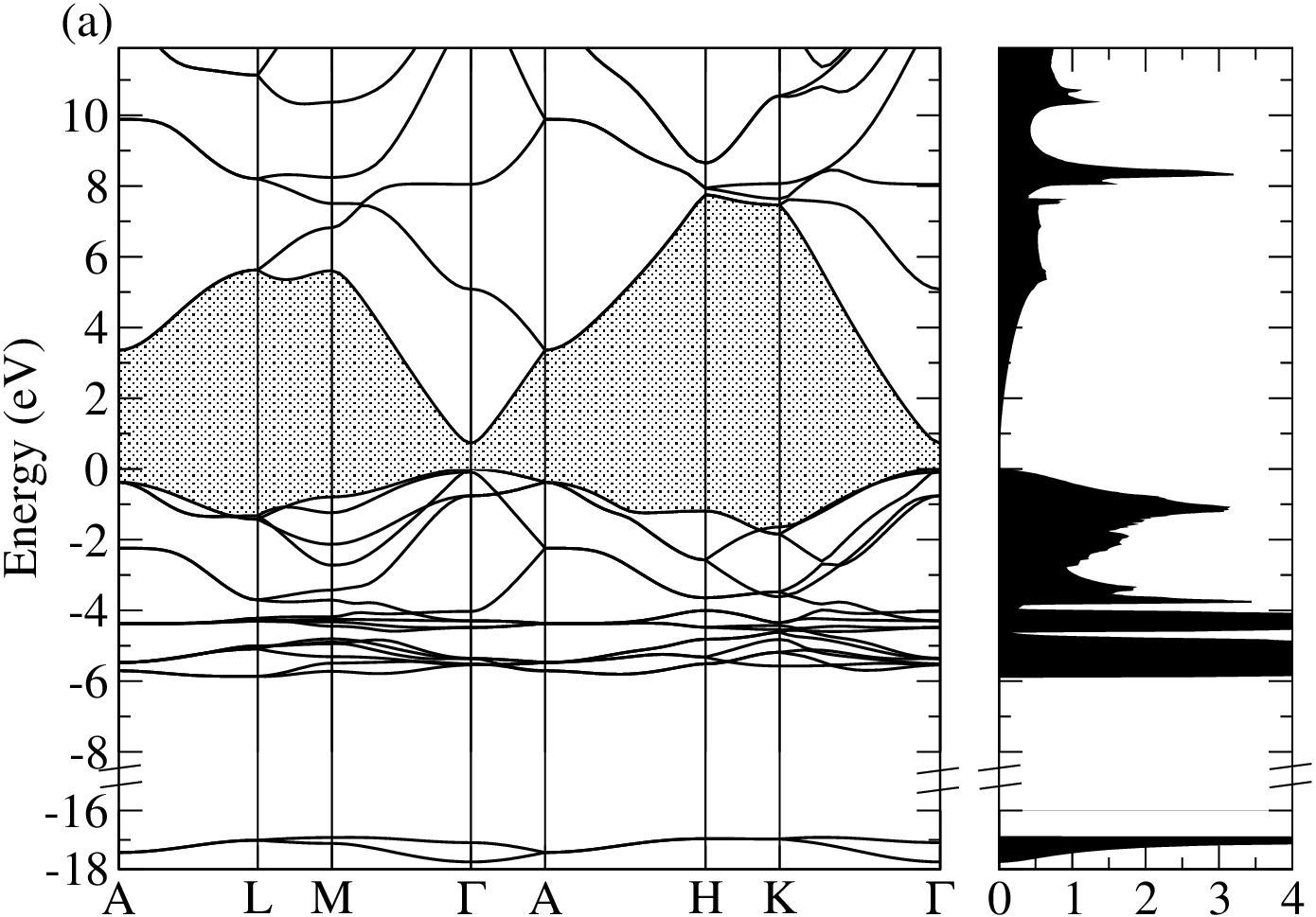}}\\
\vspace{0.1cm}
\resizebox{0.39\textwidth}{0.29\textwidth}{\includegraphics*{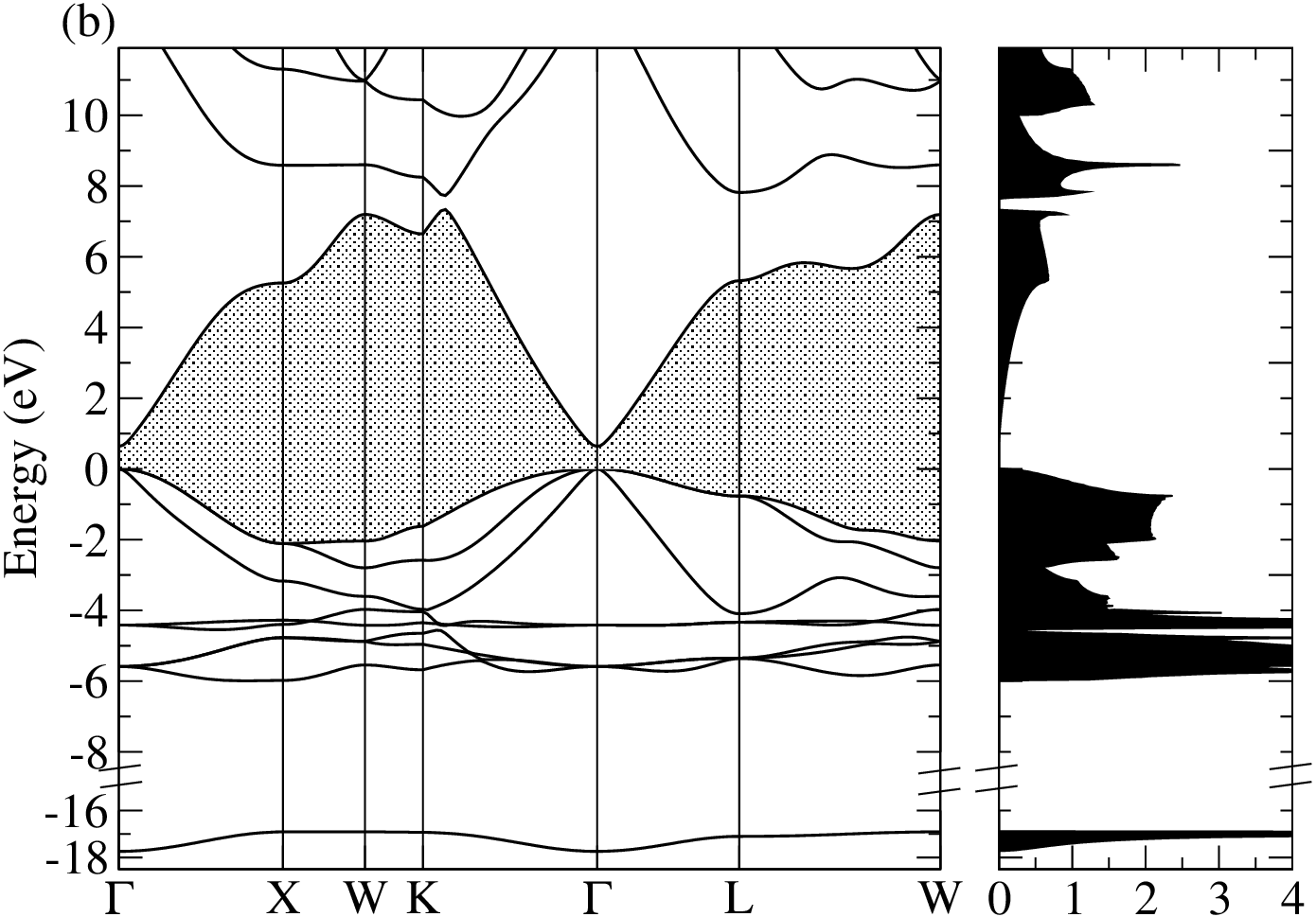}}\\
\vspace{0.1cm}
\resizebox{0.39\textwidth}{0.29\textwidth}{\includegraphics*{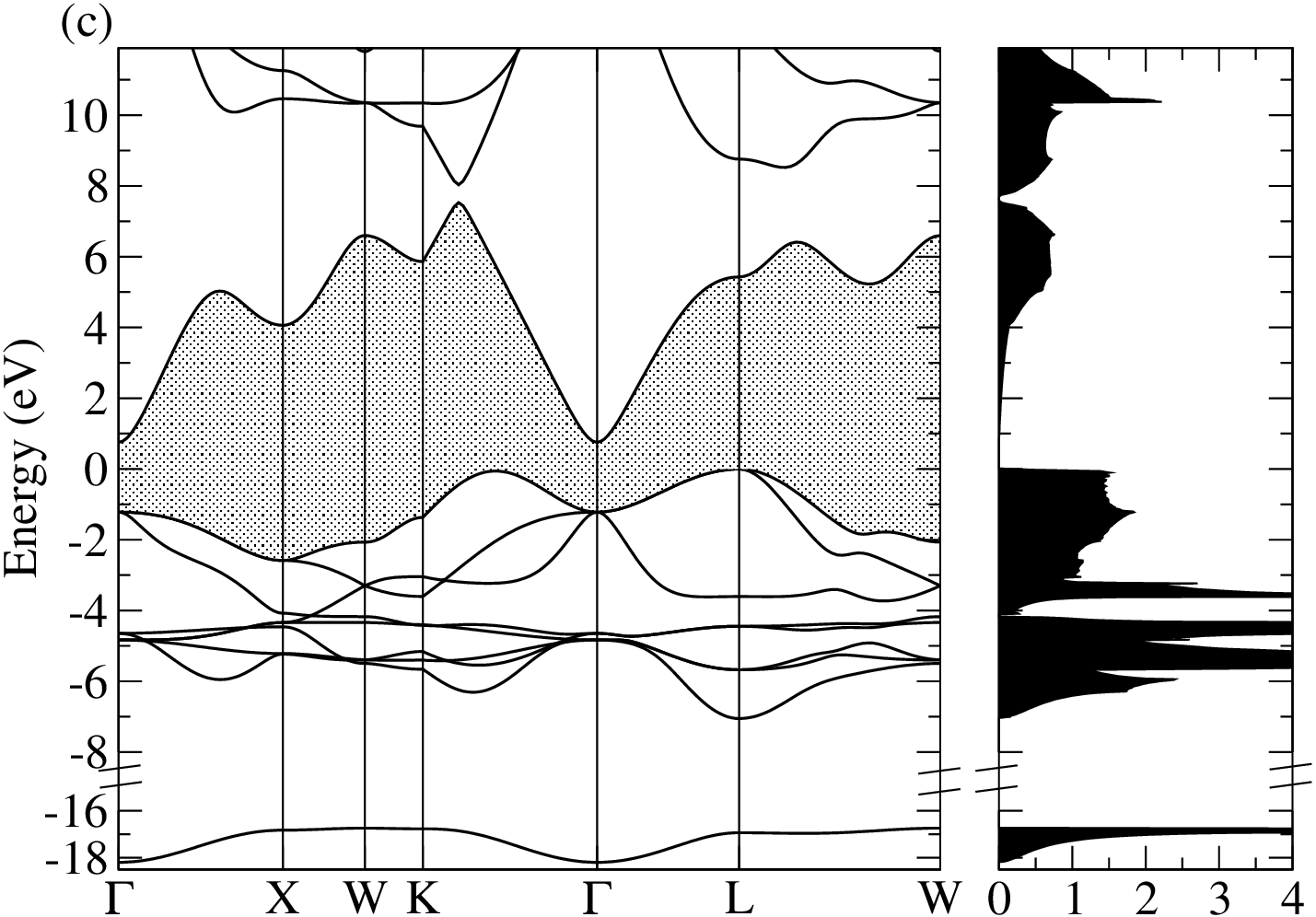}}\\
\vspace{0.1cm}
\resizebox{0.39\textwidth}{0.30\textwidth}{\includegraphics*{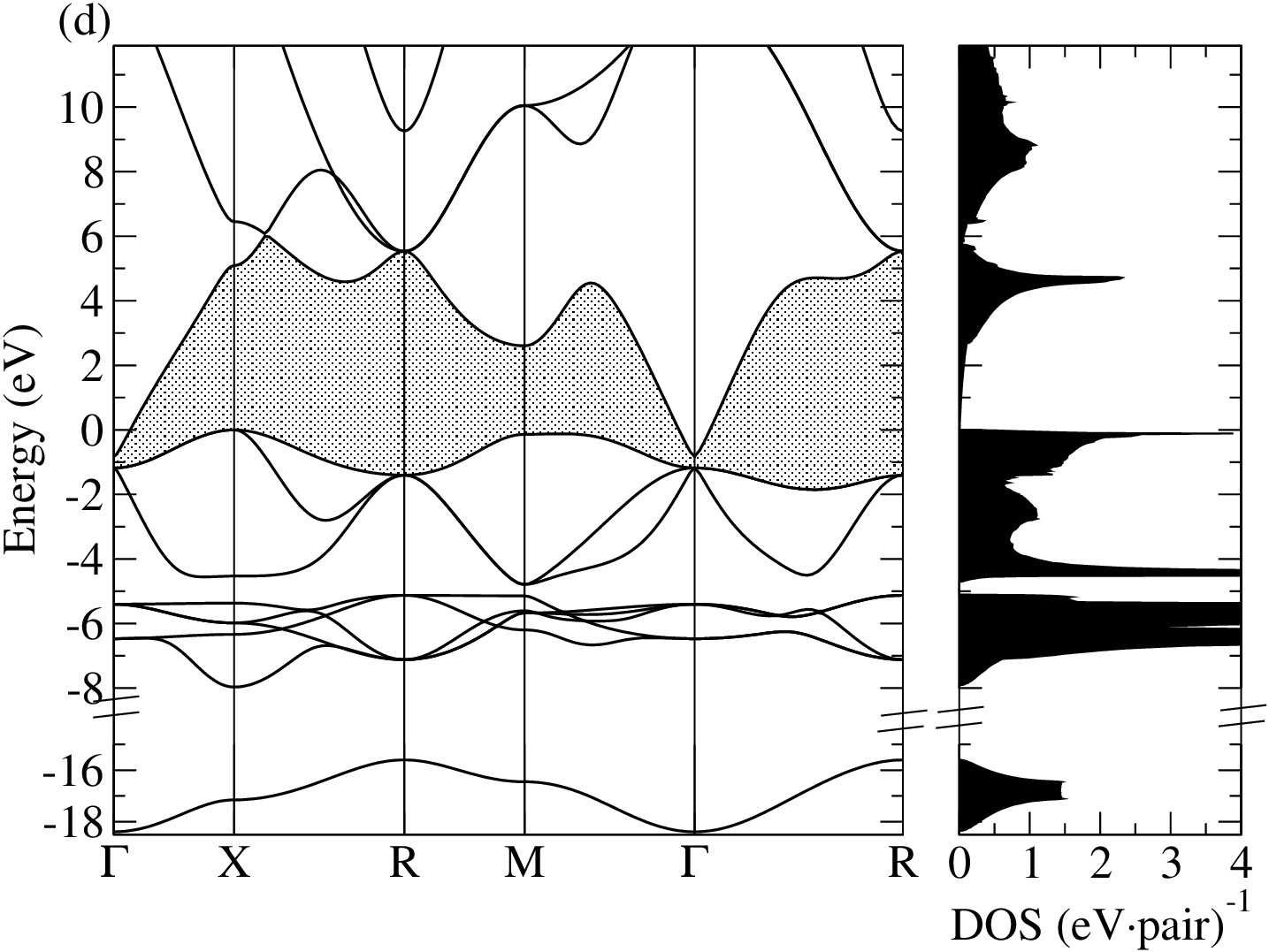}}
\caption{\label{fig:electronic_zno} Band structure and density of
  states (normalized per pair) for ZnO polymorphs 
calculated within the DFT-GGA framework: (a) wurtzite, (b)
zinc blende, (c) rocksalt, and (d) cesium chloride. The shaded region 
indicates the fundamental gap. The valence band maximum is chosen as energy zero.}
\end{figure}

\begin{figure}
\resizebox{0.39\textwidth}{0.29\textwidth}{\includegraphics*{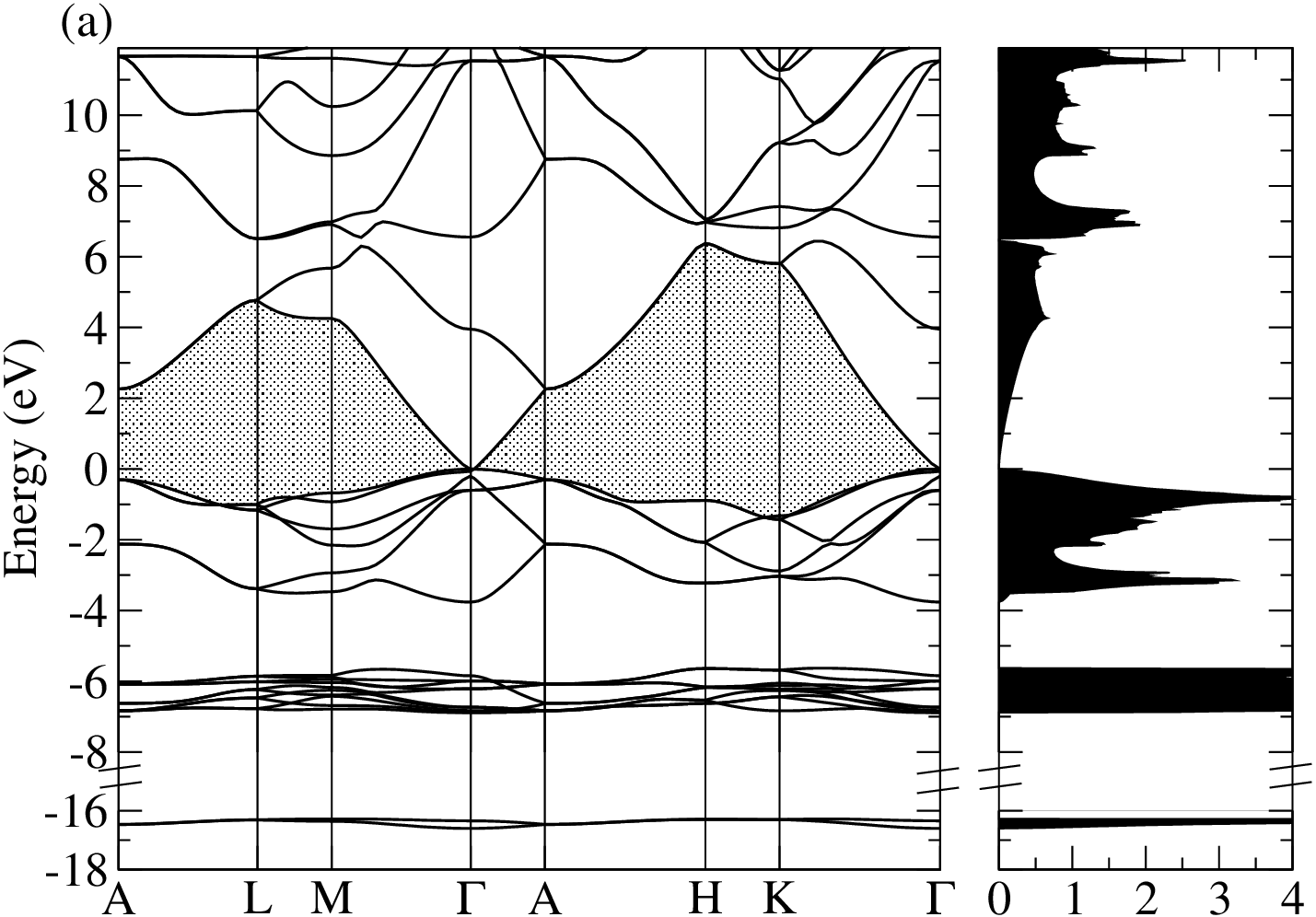}}\\
\vspace{0.1cm}
\resizebox{0.39\textwidth}{0.29\textwidth}{\includegraphics*{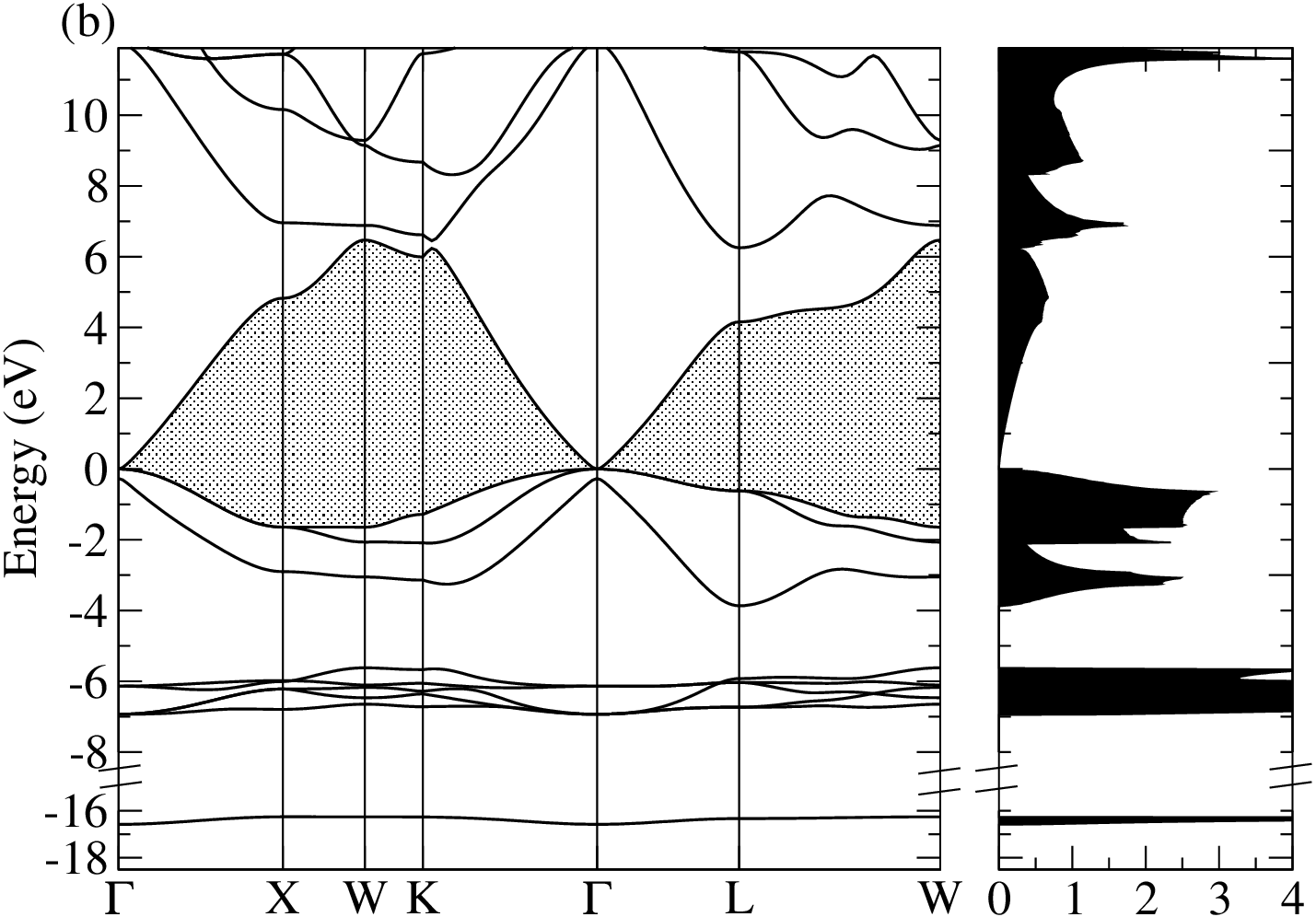}}\\
\vspace{0.1cm}
\resizebox{0.39\textwidth}{0.29\textwidth}{\includegraphics*{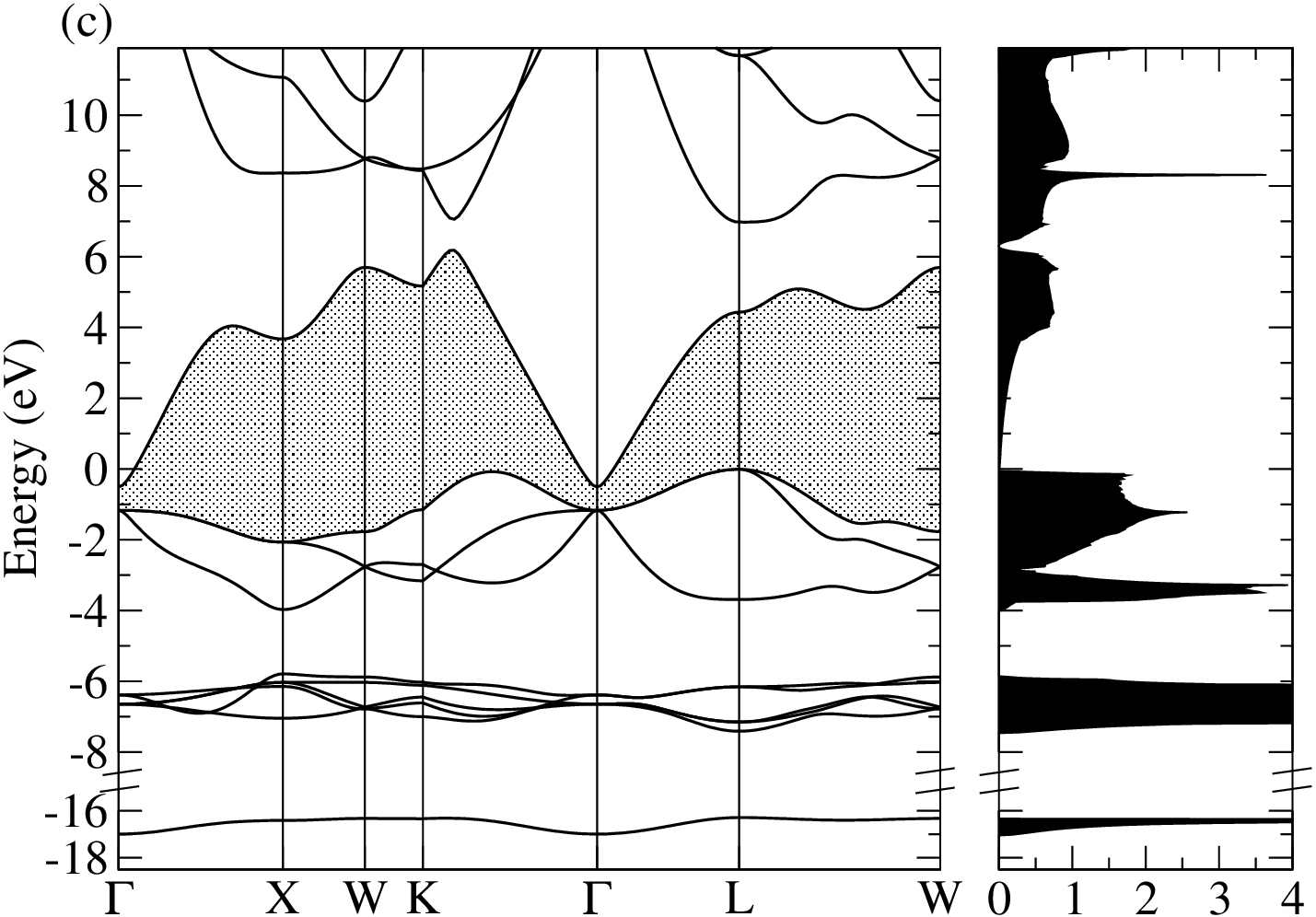}}\\
\vspace{0.1cm}
\resizebox{0.39\textwidth}{0.30\textwidth}{\includegraphics*{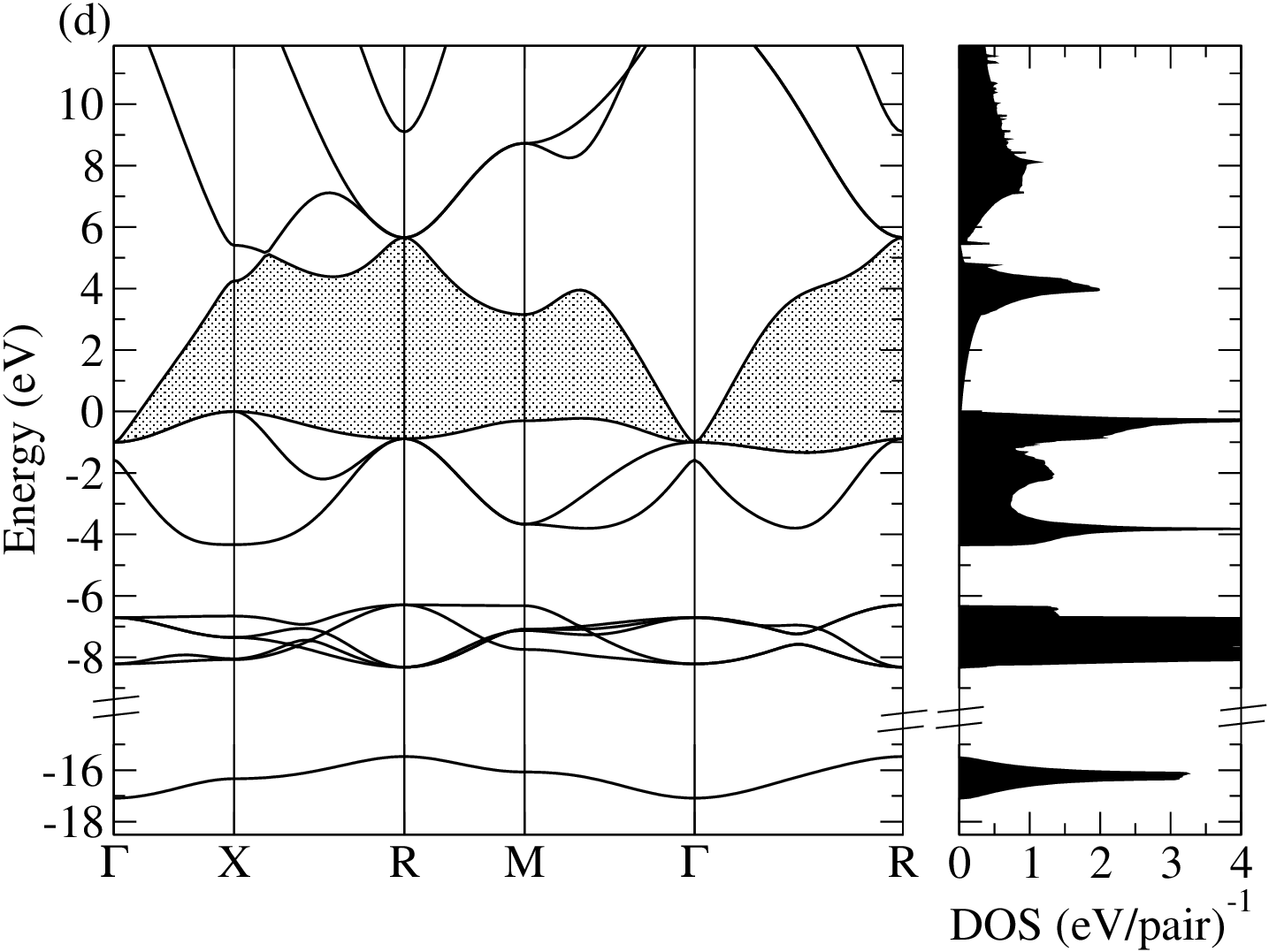}}
\caption{\label{fig:electronic_cdo} Band structure and density of
  states (normalized per pair) for CdO polymorphs 
calculated within the DFT-GGA framework: (a) wurtzite, (b)
zinc blende, (c) rocksalt, and (d) cesium chloride. The shaded region 
indicates the fundamental gap. The valence band maximum is chosen as energy zero.}
\end{figure}

To study the influence of the atomic geometry, more precisely the polymorph
of the group-II oxide compound, we show in Figs.~\ref{fig:electronic_mgo},
\ref{fig:electronic_zno}, and \ref{fig:electronic_cdo} the band structures as
calculated within DFT-GGA for the most stable polymorphs in a
certain volume range around the equilibrium volume (cf. Sec. III).
These are the rocksalt and cesium
chloride structures, supplemented by the wurtzite structure and the zinc-blende
structure.\\
By examining the electronic structure in the KS approach one
neglects the excitation aspect \cite{17,18} and, hence, underestimates
the resulting energy gaps and interband transition energies. In GW
approximation the corresponding quasiparticle (QP) energy corrections due to
the exchange-correlation self-energy \cite{17,18,19} amount to about
2.0~eV for ZnO \cite{20} and 3.6~eV for MgO \cite{21}. Nevertheless
the independent-particle approximation \cite{22} is frequently a good
starting point for the description of optical properties. 

For MgO the band structures and densities of states in
Fig.~\ref{fig:electronic_mgo} indicate an insulator or wide-gap
semiconductor independent of the polymorph. Except the
CsCl-structure, which has an indirect gap between $M$ and $\Gamma$,
they all have direct fundamental gaps at the $\Gamma$ point in the
BZ. We find rather similar values
4.5~eV ($rs$), 3.5~eV ($zb$), 4.2~eV (NiAs), and 3.3~eV ($h$-MgO) for the gaps of cubic
and hexagonal crystals despite the different lattice constants,
coordination, and bonding. There are also similarities in the atomic
origin of the bands. The
O2$s$ states give rise to weakly dispersive bands 15--18~eV
below the valence-band maximum (VBM). They are separated by an
ionic gap of 10--12~eV from the uppermost valence bands with band
widths below 5~eV. Due to the high ionicity of the bonds the
corresponding eigenstates predominately possess O2$p$ character. For
the same reason the lowest conduction bands can be traced back to
Mg3$s$ states. Also mentionable is the relatively weak dispersion
of the uppermost valence bands for the
wurtzite and zinc-blende structures which leads, compared with the
rocksalt and CsCl structures, to a quite high DOS near the VBM.\\
For the equilibrium rocksalt polymorph we calculate besides the direct gap
at $\Gamma$ other direct gaps at $X$ and $L$ in the BZ:
$E_g(X)=10.43$~eV and $E_g(L)=8.37$~eV. These values are in good
agreement with results of other DFT calculations using an LDA
exchange-correlation functional and a smaller lattice constant \cite{21,45,46}. In their
works \cite{21}/\cite{45} these authors computed QP
gap openings of about 3.59/2.5~eV ($\Gamma$), 3.96/2.5~eV ($X$), and
4.00/2.5~eV ($L$) in rough agreement with the values 3.06~eV of Shirley
\cite{46} and 2.94~eV derived from the simple Bechstedt-DelSole
formula for tetrahedrally bonded crystals \cite{47}. With these
values one obtains QP gaps which approach the experimental values
of $E_g(\Gamma)=7.7$~eV, $E_g(X)=13.3$~eV, and $E_g(L)=10.8$~eV
\cite{6} or $E_g(\Gamma)=7.83$~eV \cite{65}. 
A more sophisticated QP
value for the fundamental gap is $E_g(\Gamma)=7.79$~eV \cite{66}.

The electronic structures of ZnO plotted in
Fig.~\ref{fig:electronic_zno} show several similarities to those of
MgO. However, the O2$s$ bands now appear 
at about 15.5--18.5~eV below the VBM, and the uppermost valence bands of
predominantly O2$p$ character are found in the range from 0 to $-4$~eV. The
lowest conduction-band states (at least near $\Gamma$) are
dominated by Zn4$s$ states. In comparison with the fundamental energy
gaps of MgO those of ZnO are smaller. We compute 0.73~eV ($w$),
0.64~eV ($zb$), 1.97~eV ($rs$) for the direct gap at
$\Gamma$. However, for the rocksalt geometry the VBM occurs at the $L$
point and therefore this high-pressure phase is an indirect semiconductor
with a gap of $E_g(\Gamma-L)=0.75$~eV. This observation is in agreement with
findings by room-temperature absorption measurements and DFT-LDA calculations
\cite{20,31,48}. Another local valence band maximum which is almost as high as 
the one at $L$ occurs at the $\Sigma$ line between $K$ and
$\Gamma$.\\
New features, that are not observable for MgO are caused by the Zn3$d$
states. These shallow core states give rise to two groups of bands (at
$\Gamma$) clearly visible in the energy range of 4--6~eV below the
VBM, which generally show a splitting and a wave-vector dispersion
outside $\Gamma$. The huge peaks caused by these basically
Zn3$d$-derived bands are clearly visible in the DOS. Furthermore, the
Zn3$d$ states act more subtle on the band 
structure via the repulsion of $p$ and $d$ bands caused by the hybridization
of the respective states. The effects of this $pd$ repulsion can be discussed
most easily for the $\Gamma$-point in the BZ:
In the case of the $zb$ polymorph (with tetrahedral coordination and $T_d^2$ symmetry),
the hybridized anion $p(t_2)$ and cation $d(t_2)$ levels give rise to the threefold
degenerate $\Gamma_{15}(pd)$ and $\Gamma_{15}(dp)$ levels \cite{49}.
In the case of wurtzite the levels are doubled at $\Gamma$ (with
respect to the $zb$-polymorph) due to the band folding along the
[111]/[0001] direction. In addition, these states are influenced by a
crystal-field splitting. Moreover, in the case of the tetrahedrally
coordinated $zb$ and $w$ polymorphs the $pd$ repulsion reduces the
fundamental direct gaps at $\Gamma$ as the $\Gamma_{15}(pd)$ are
pushed to higher energies, while the $\Gamma_{1c}$ conduction band
minimum (CBM) remains unaffected.
In the case of the $rs$ polymorph (with sixfold coordination and octahedral O$^5_h$
symmetry), the anion $p(t_1)$ level gives rise to the threefold degenerate
$\Gamma_{15}(p)$ valence band, while the symmetry-adapted $e$ (twofold degenerate)
and $t_2$ (threefold degenerate) combination of the cation $d$ states generates
the $\Gamma_{12}(d)$ and $\Gamma_{25'}(d)$ bands. The respective states do not
hybridize and the $pd$ repulsion as well as the corresponding gap shrinkage vanishes \cite{49}.
However in other regions of the BZ, the bands are subject to the $pd$ repulsion
and thereby raised at points away from $\Gamma$.
Consequently the $pd$ repulsion, or more exactly its symmetry forbiddance at
$\Gamma$ is the reason why the $rs$ polymorph is an indirect semiconductor.
Compared to InN the $pd$ repulsion is larger but does not give rise to a
negative $\Gamma_1(s)$--$\Gamma_{15}(pd)$ gap as found there \cite{67}.\\ 
Using QP corrections the small energy gaps obtained within DFT-GGA are
significantly opened. With appropriate parameters as bond
polarizability 0.78, nearest-neighbor distance 2.01~{\AA}, and
electronic dielectric constant 4.0 \cite{51}) for the
tetrahedrally coordinated ZnO the Bechstedt-DelSole formula gives
a QP shift of about 1.95~eV. This gap correction is bracketed by
other approximate values of 0.85~eV \cite{52}, 1.04~eV \cite{69}, and 2.49~eV \cite{53}.
However, it agrees well with the value 1.67~eV derived within a
sophisticated QP calculation \cite{20}. Our values
resulting for the direct fundamental gap at $\Gamma$ 2.68~eV ($w$)
or 2.59~eV ($zb$) are larger than the QP gap value of
2.44~eV \cite{20} but still clearly underestimate the
experimental gap of 3.44~eV \cite{54,55}. This underestimation is
sometimes related to the overscreening within the random-phase
approximation (RPA) used in the QP approach \cite{20}.
We claim that one important reason is the overestimation of
the $pd$ repulsion and, hence, the too high position of the VBM in
energy due to the too shallow $d$ bands in LDA/GGA.
Even when including many-body effects, the Zn3$d$ bands are
still to high in energy. For a better treatment of the
QP effects also for the semicore $d$ states, the $pd$
repulsion should be reduced, which requires the inclusion of non-diagonal elements
of the self-energy operator or the use of a starting point different from
LDA/GGA, e.g. a generalized KS scheme \cite{72}. A recent work \cite{48} reported that
rocksalt ZnO has an indirect band gap of 2.45$\pm$0.15~eV measured
from optical absorption. Using the above estimated QP shift
of 1.95~eV and the indirect gap in DFT-LDA quality of 0.75~eV one
finds a QP value of 2.60~eV close to the measured
absorption edge.

In the case of the same polymorph the band structure and density
of states of CdO in Fig.~\ref{fig:electronic_cdo} shows several
similarities with those for ZnO. At about 15.5--17~eV below the VBM occur the
O2$s$ bands, and the Cd4$d$ bands are observed in the energy
interval from $-5.9$ to $-6.6$~eV. The uppermost O2$p$-derived valence bands
possess a maximum band width of about 3.3~eV, a value smaller than
that found in recent photoemission studies \cite{56}. Usually the
conduction bands are well separated from the valence
bands. However, for all polymorphs the lowest conduction band shows a
pronounced minimum at the center of the BZ. Within an energy range of
a few tenths of an 
electron volt above its bottom, this band is isotropic but highly
nonparabolic.\\
In $rs$-CdO we find a direct gap of about 0.66~eV at the $\Gamma$ point.
As in the case of $rs$-ZnO also for $rs$-CdO the maxima of the valence bands
occur at the $L$ point and at the $\Sigma$ line between
$\Gamma$ and $K$. These maxima lie above the CBM, which results in negative
indirect gaps of about $-0.51$ and $-0.43$~eV, and therefore our band structure
indicates a halfmetal. As described for $rs$-ZnO, the $pd$
repulsion is responsible for this effect \cite{68}.
Our band structure is in qualitative agreement with a DFT-LDA
calculation \cite{56}, in particular with respect to the band
dispersions. However, Ref. \cite{56} found positive direct and
indirect gap values. The gap opening of about 1~eV with
respect to our values may be a consequence of the used basis set
restricted to a few Gaussians. Corresponding experimental
values are 0.84 and 1.09~eV for the indirect gaps and 2.28~eV for
the lowest direct gap at $\Gamma$. Their comparison with the
values calculated within DFT-GGA indicates effective QP
gap openings of about 1.3--1.7~eV. However, these values should
be considerably influenced by the $pd$ repulsion, at least outside
the BZ center. While the Cd4$d$ bands are about 6.4 or 6.6~eV below
the valence band maximum at $\Gamma$, the experimental distances are
about 12.4 or 13.3~eV (with respect to the Fermi level) \cite{57}.
A more recent measurement indicates an average binding energy of
Cd4$d$ relative to VBM of about 9.4~eV \cite{56}.
Altogether the band structures of CdO are rather
similar to those of InN \cite{23}. Within DFT-GGA in both cases small
negative $s$-$p$ fundamental gaps are found near $\Gamma$. The
main difference is related with the position of the 4$d$ bands,
because the binding energy of In4$d$ electrons is larger than that of
Cd4$d$ electrons.

The band structures presented in Figs.~\ref{fig:electronic_mgo},
\ref{fig:electronic_zno}, and~\ref{fig:electronic_cdo} show clear 
chemical trends along the series MgO, ZnO, and CdO for a fixed
crystal structure. To clarify these trends we list in
Table~\ref{tab:trans} some energy positions for the two most important
polymorphs -- rocksalt and wurtzite. Among these
energies we also list the lowest conduction and highest valence bands
for two high-symmetry points in the corresponding fcc ($\Gamma$,
$L$) or hexagonal ($\Gamma$, $A$) BZ. In the case of the cubic systems
the position of the twofold (threefold) degenerate
shallow core $d$ level $\Gamma_{12}$ ($\Gamma_{25'}$) at $\Gamma$
is given. For the rocksalt phase as well as the
wurtzite polymorph the level positions follow a clear chemical
trend with the cations Mg, Zn, and Cd. The energetical
position of the empty conduction-band levels as well as the
filled valence-band states decreases with respect to the VBM.
Consequently, the average gaps decrease along the series MgO, ZnO,
and CdO. At least for tetrahedrally coordinated compounds the general
trend is governed by both, the splitting of the $s$- and
$p$-valence energies as well as the nearest-neighbor distances
\cite{51}. In the case of wurtzite the effect of the non-monotonous
behaviour of both quantities may give a monotonous net effect.
Another atomic tendency, the increasing cation-$p$--anion-$d$
splitting from ZnO to CdO \cite{49,51}, is correlated with the
increase of the distance of the $d$ bands to the VBM.\\
We have also computed the volume deformation potentials for the direct gaps
at $\Gamma$ in rocksalt MgO to $a_V=-9.39$~eV and in wurtzite ZnO to
$a_V=-1.55$~eV, as well as for the indirect gap between $L$ and $\Gamma$
in rocksalt CdO to $a_V=-1.87$~eV. The absolute value of the gap deformation
potential of MgO is clearly larger than those of ZnO or CdO, due to the stronger bonding in MgO.
Despite different gap states and ground-state polymorphs the values for ZnO and CdO are rather similar.
Their smallness is a consequence of the large ionicity and the relatively large bond length. 
According to the estimate of \cite{8}, this may explain the small gaps of ZnO and CdO
in comparison to ZnS and CdS. The deformation potential $a_V=-1.55$~eV calculated for
$w$-ZnO within DFT-GGA is much smaller compared to the experimental value of $a_V=-3.51$~eV \cite{70}.
The reason should be the neglect of the large QP corrections not taken into account.

\begin{table}
\caption{\label{tab:trans} Characteristic energy levels (in eV) in the band
structure of the rocksalt and wurtzite polymorphs of MgO, ZnO, and
CdO as calculated within the DFT-GGA framework. The center
$\Gamma$ of the BZ and an $L$ and $A$ point at the BZ surface in
[111]/[0001] direction are chosen for the Bloch wave vectors. The
lowest conduction ($c$) bands and the highest valence ($v$) bands are
studied. In the case of rocksalt also the positions of the
$\Gamma_{12}$ and $\Gamma_{25'}$ $d$ bands are given. The
uppermost $\Gamma_{15v}$ or $\Gamma_{6v}$ valence band at $\Gamma$
is used as energy zero. }
\begin{ruledtabular}
\begin{tabular}{|c|c|c|c|c|c|c|c|}
\multicolumn{4}{|c|}{rocksalt} & \multicolumn{4}{c|}{wurtzite}\\
\hline
 level & MgO & ZnO & CdO & level & MgO & ZnO & CdO \\ \hline
 & & & & $\Gamma_{3c}$ & 8.62 & 5.08 & 3.95\\
  $\Gamma_{1c}$ & 4.50 & 1.97 & 0.66 & $\Gamma_{1c}$ & 3.68 &
  0.73 & $-0.20$ \\
 $\Gamma_{15v}$ & 0.00 & 0.00 & 0.00 & $\Gamma_{6v}$ & 0.00 & 0.00 & 0.00 \\
 & & & & $\Gamma_{1v}$ & 0.31 & $-0.09$ & $-0.07$ \\
 $\Gamma_{12}$ & -- & $-3.43$ & $-5.22$ & $\Gamma_{5v}$ & $-0.30$ & $-0.76$ & $-0.60$ \\
 $\Gamma_{25'}$ & -- & $-3.61$ & $-5.48$ & $\Gamma_{3v}$ & $-2.79$ & $-4.03$ & $-3.76$ \\
 $L_{2'c}$ & 7.76 & 6.65 & 5.60 & $A_{1,3c}$ & 6.11 & 3.36 & 2.26 \\
 $L_{3v}$ & $-0.62$ & 1.22 & 1.17 & $A_{5,6v}$ & $-0.15$ & $-0.37$ & $-0.27$  \\
 $L_{1v}$ & $-4.48$ & $-2.38$ & $-2.52$ & $A_{5v}$ & $-1.61$ & $-2.24$ & $-2.12$ \\ \hline
 \end{tabular}
\end{ruledtabular}
\end{table}

\subsection{Dielectric functions and optical properties}

\begin{figure}
\resizebox{0.45\textwidth}{!}{\includegraphics*{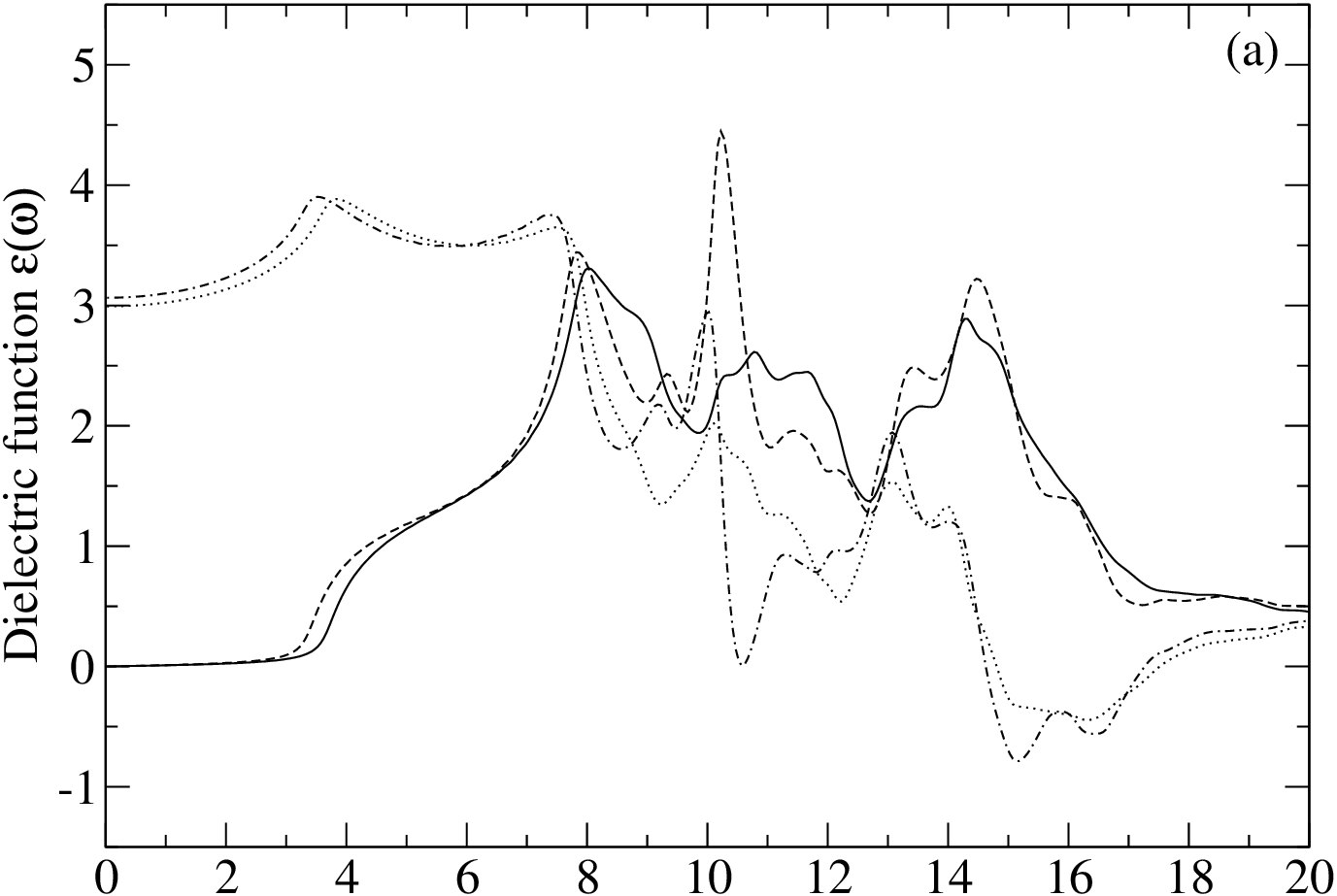}}\\
\resizebox{0.45\textwidth}{!}{\includegraphics*{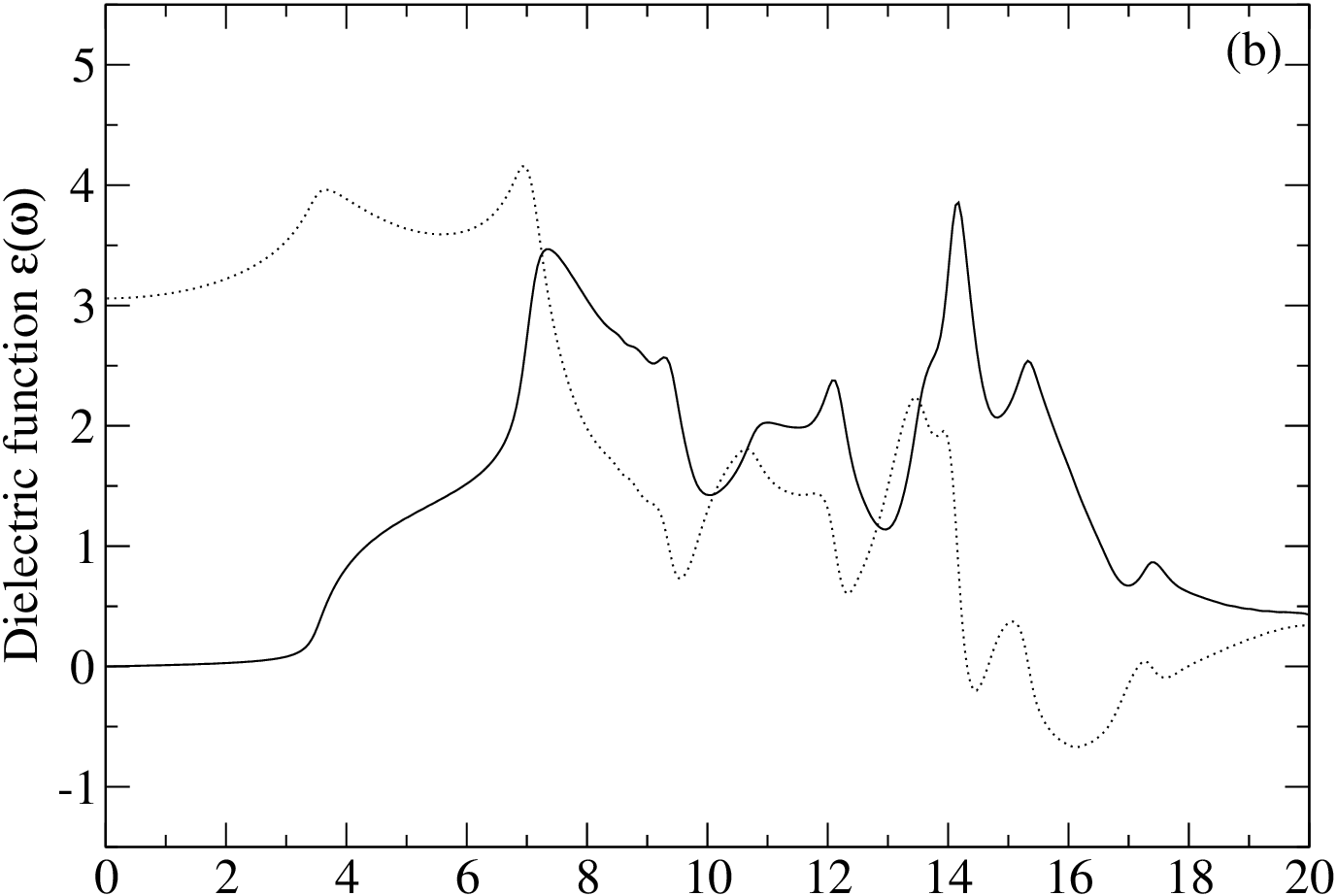}}\\
\resizebox{0.45\textwidth}{!}{\includegraphics*{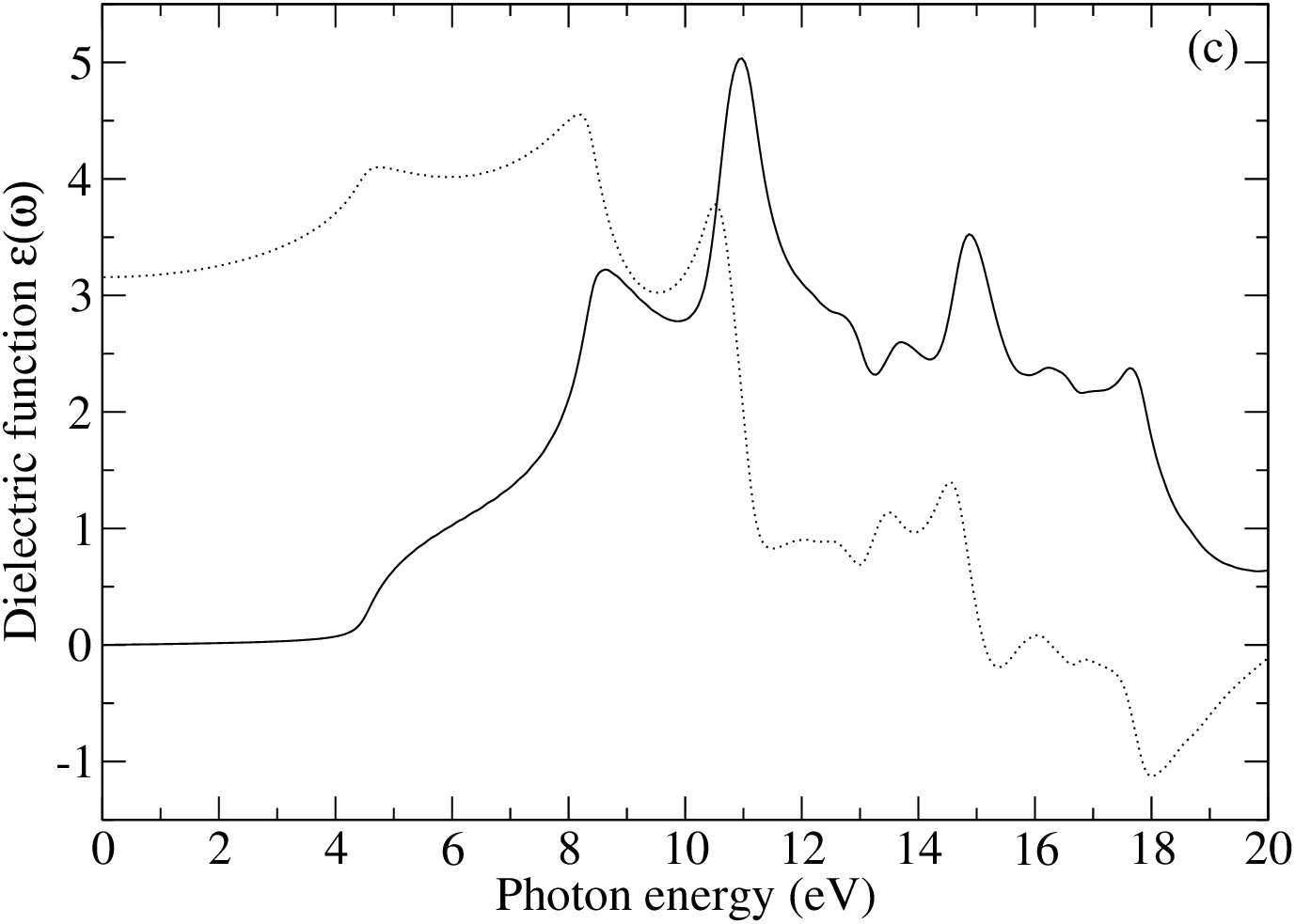}}
\caption{\label{fig:optic_mgo} Real and imaginary part of the
  frequency-dependent dielectric function for MgO polymorphs wurtzite
  (a), zinc blende (b), and rocksalt (c) as calculated within the
independent-particle approximation using Kohn-Sham eigenstates and
eigenvalues from DFT-GGA. Imaginary part: solid line, real part:
dotted line. In the case of wurtzite besides the tensor components
$\varepsilon_{xx}(\omega)=\varepsilon_{yy}(\omega)$ also the
$zz$-component $\varepsilon_{zz}(\omega)$ is presented (imaginary
part: dashed line, real part: dash-dotted line).}
\end{figure}

\begin{figure}
\resizebox{0.45\textwidth}{!}{\includegraphics*{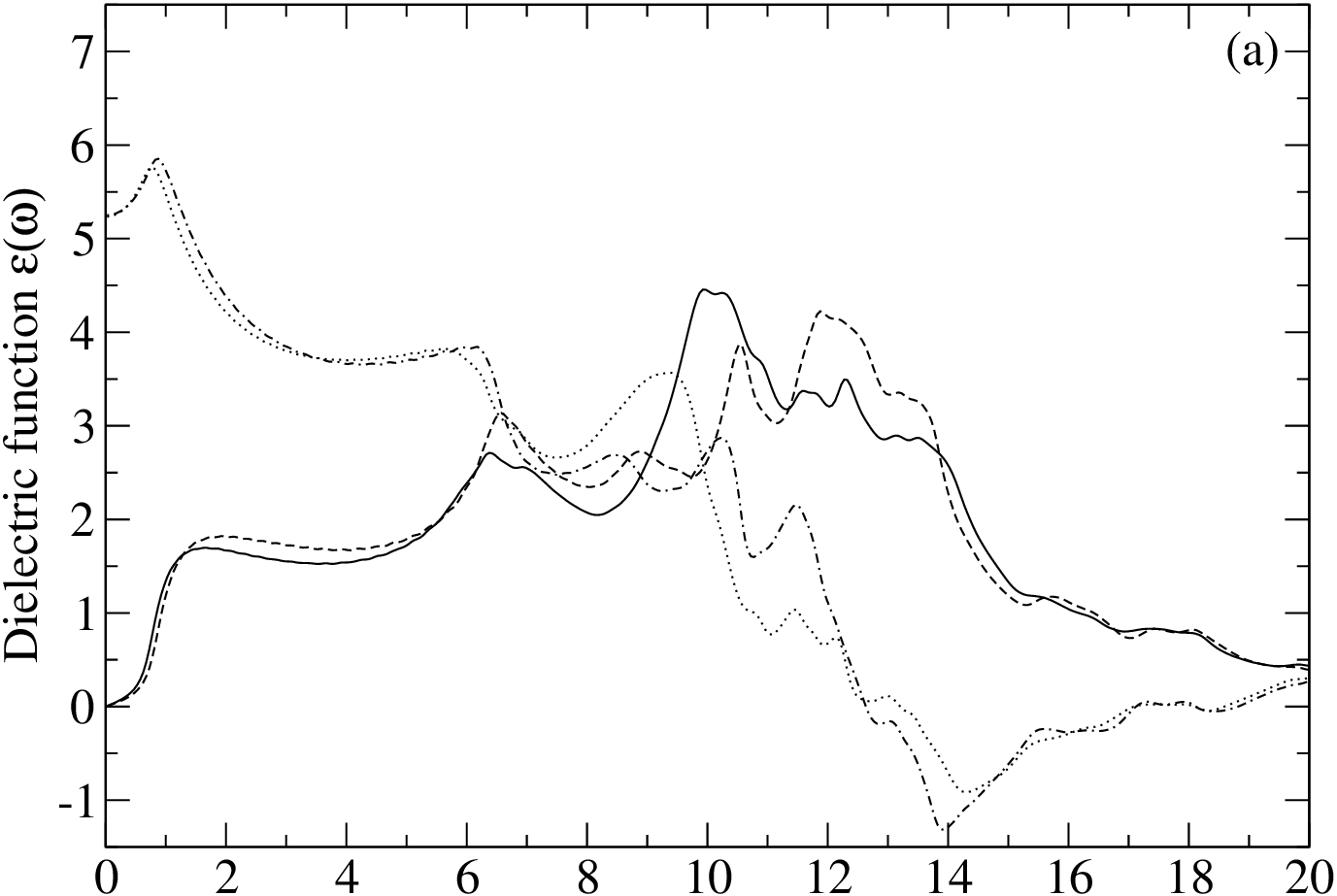}}\\
\resizebox{0.45\textwidth}{!}{\includegraphics*{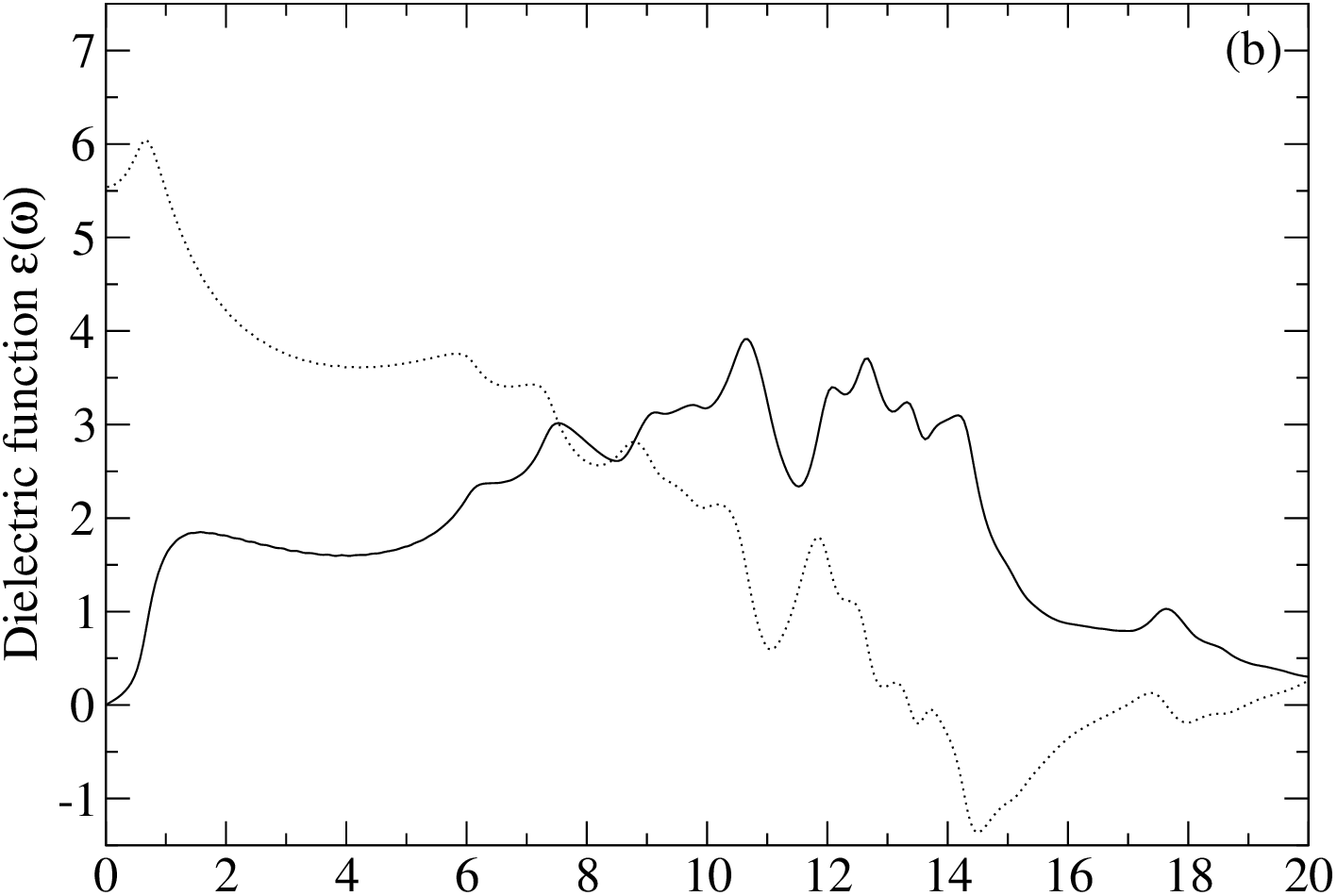}}\\
\resizebox{0.45\textwidth}{!}{\includegraphics*{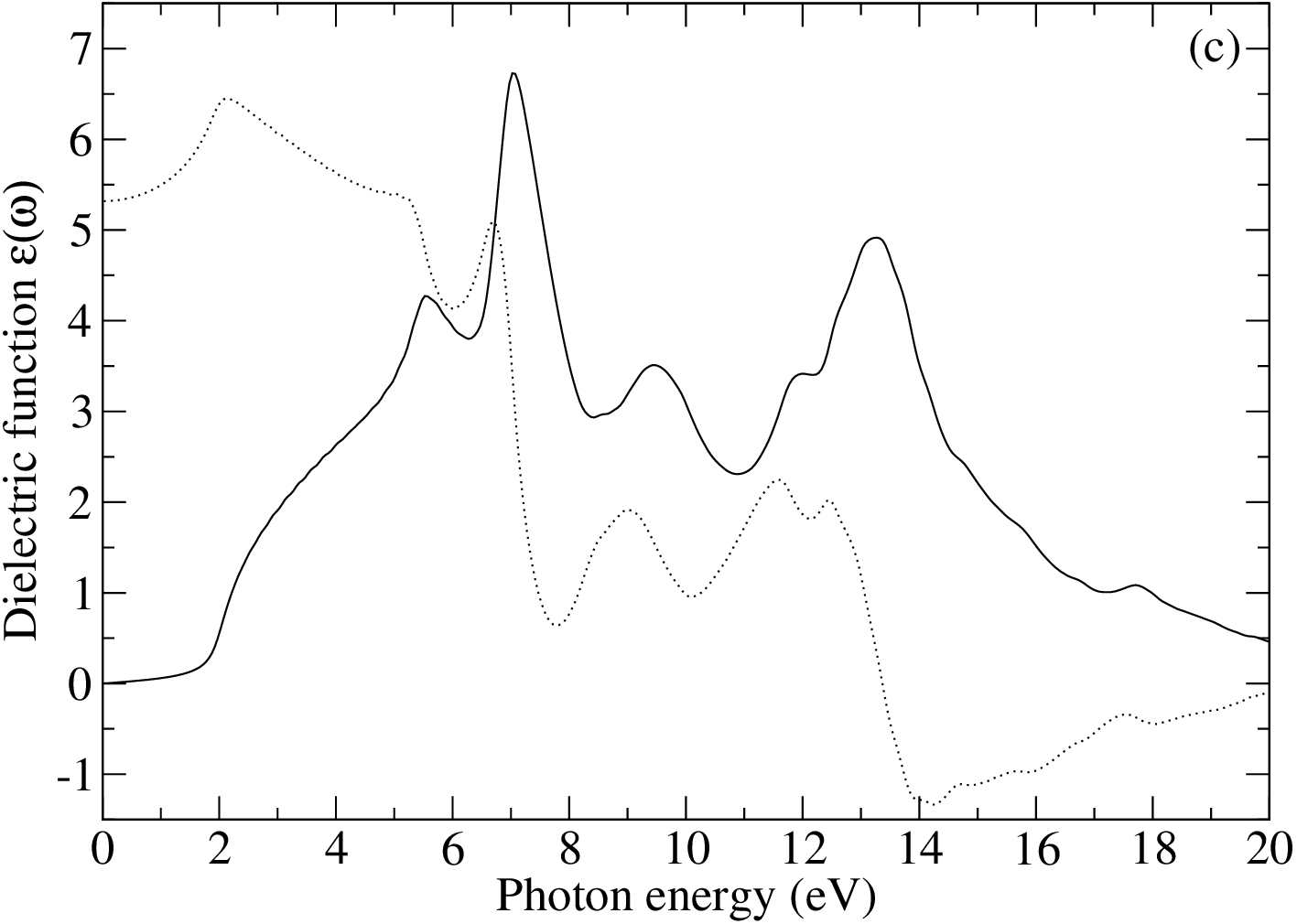}}
\caption{\label{fig:optic_zno} Real and imaginary part of the frequency-dependent
dielectric function for ZnO polymorphs wurtzite (a), zinc blende
(b), and rocksalt (c) as calculated within the
independent-particle approximation using Kohn-Sham eigenstates and
eigenvalues from DFT-GGA. Imaginary part: solid line, real part:
dotted line. In the case of wurtzite besides the tensor components
$\varepsilon_{xx}(\omega)=\varepsilon_{yy}(\omega)$ also the
$zz$-component $\varepsilon_{zz}(\omega)$ is presented (imaginary
part: dashed line, real part: dash-dotted line). }
\end{figure}

\begin{figure}
\resizebox{0.45\textwidth}{!}{\includegraphics*{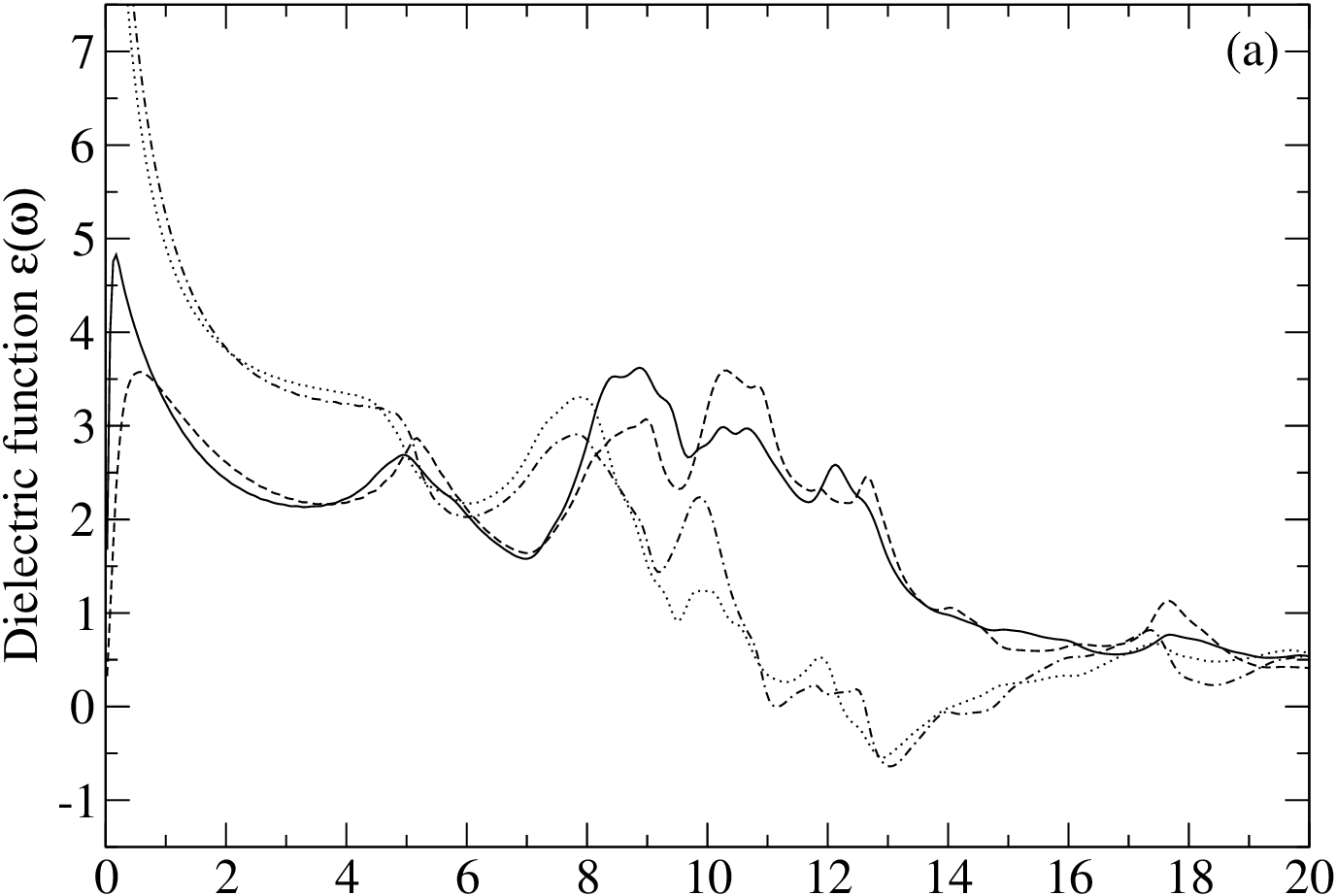}}\\
\resizebox{0.45\textwidth}{!}{\includegraphics*{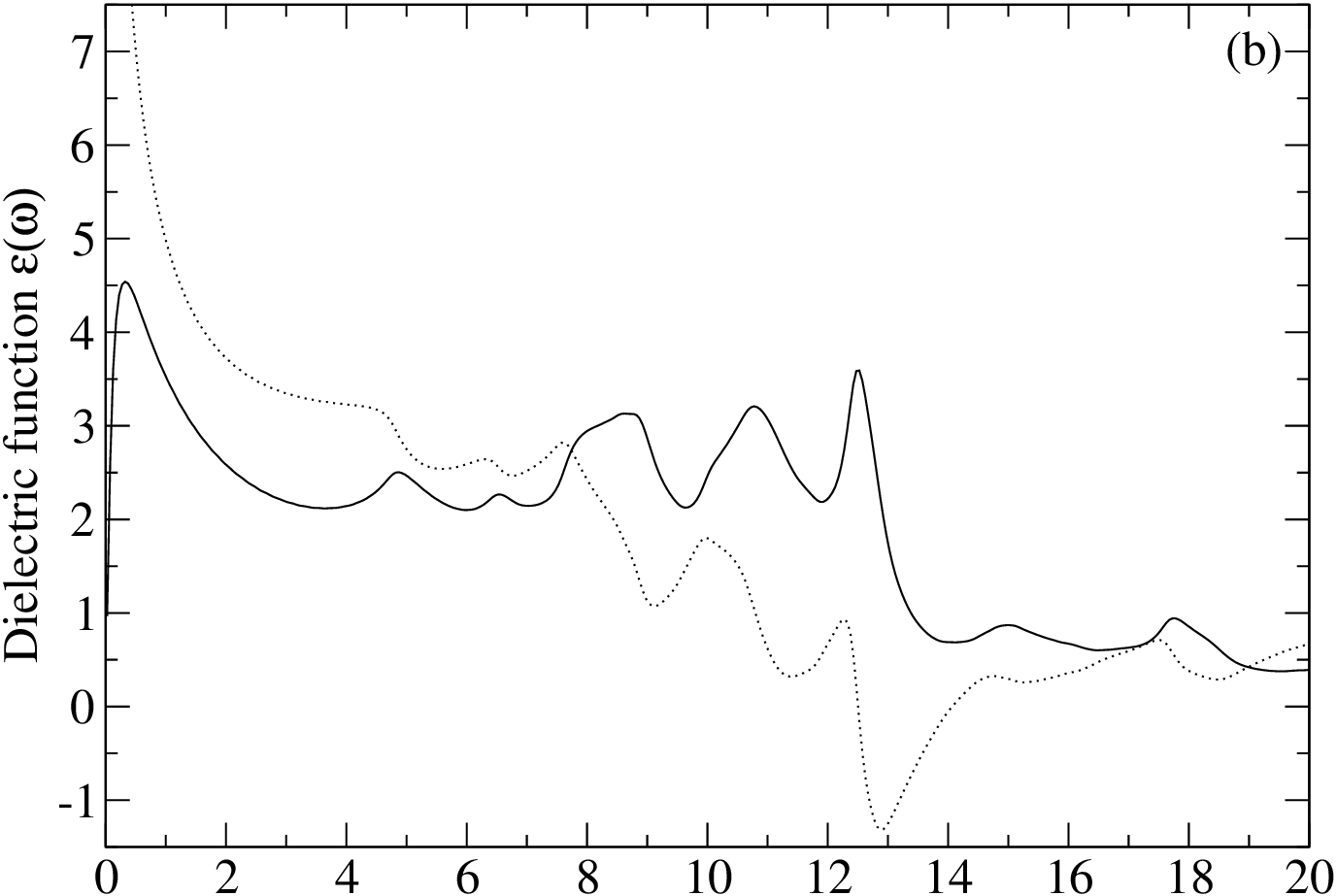}}\\
\resizebox{0.45\textwidth}{!}{\includegraphics*{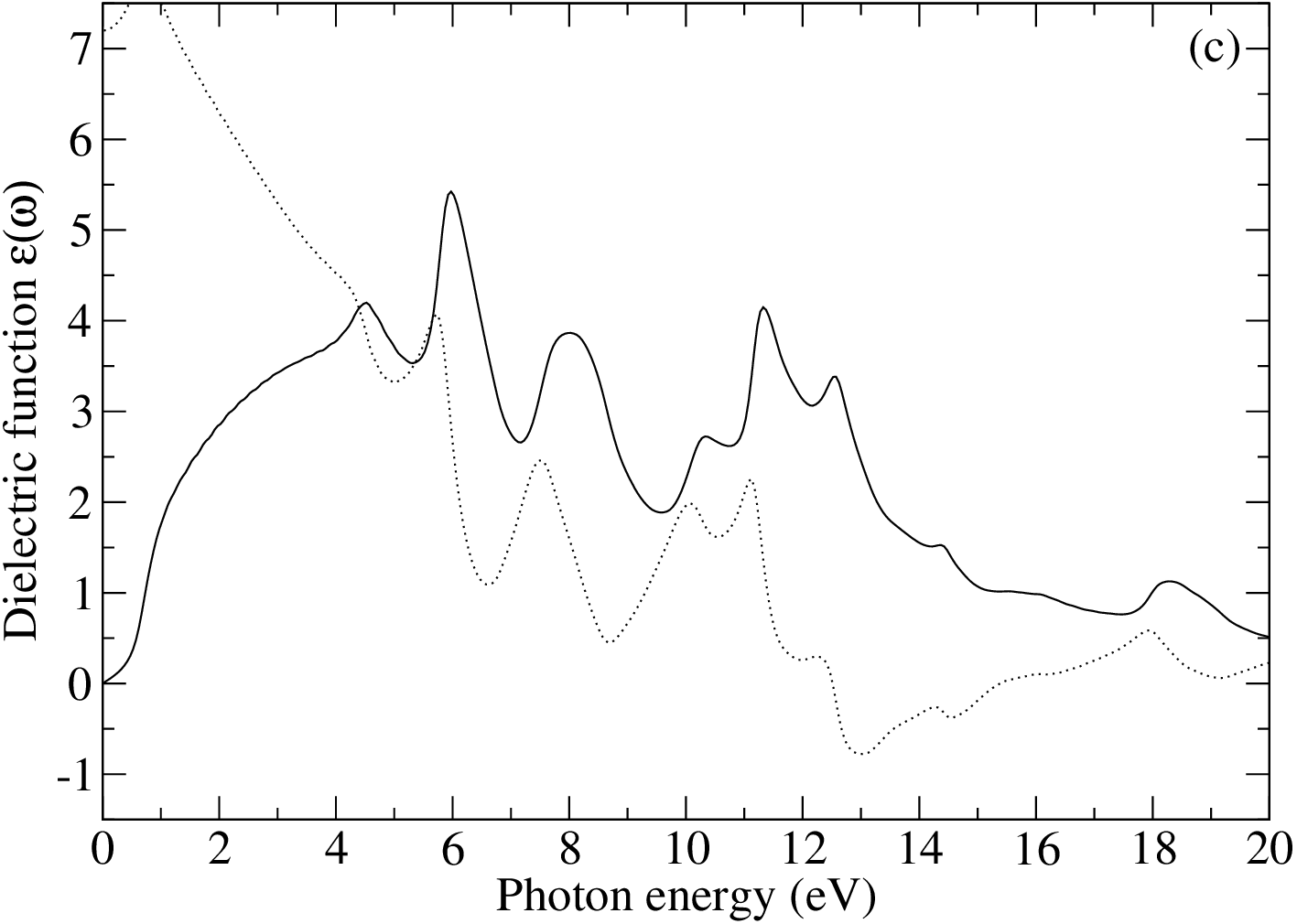}}
\caption{\label{fig:optic_cdo} Real and imaginary part of the
  frequency-dependent dielectric function for CdO polymorphs wurtzite
  (a), zinc blende (b), and rocksalt (c) as calculated within the
independent-particle approximation using Kohn-Sham eigenstates and
eigenvalues from DFT-GGA. Imaginary part: solid line, real part:
dotted line. In the case of wurtzite besides the tensor components
$\varepsilon_{xx}(\omega)=\varepsilon_{yy}(\omega)$ also the
$zz$-component $\varepsilon_{zz}(\omega)$ is presented (imaginary
part: dashed line, real part: dash-dotted line).  }
\end{figure}

For computing the dielectric function $\varepsilon(\omega)$ we
have chosen the wurtzite and rocksalt polymorphs since they
give the equilibrium geometries. In addition, we show spectra
for zinc-blende crystals because of the mentioned similarity with
wurtzite, but without its optical anisotropy.
In Figs.~\ref{fig:optic_mgo}, \ref{fig:optic_zno}, and
\ref{fig:optic_cdo} the influence of the crystallographic
structure and of the anion on the dielectric function
of a group-II oxide is demonstrated.\\
We observe that the change of the coordination of
the atoms has a strong influence, which perhaps is best noticeable
in the imaginary parts of the dielectric function. Going from
the fourfold ($w$, $zb$) to the sixfold ($rs$) coordination,
the absorption edge shifts towards higher photon energies and the oscillator
strength increases. Thereby the screening sum rule is less
influenced.
One observes a clear increase of the (high-frequency) electronic
dielectric constant $\varepsilon_\infty=\Re\left(\varepsilon(0)\right)$ along the
series MgO, ZnO, and CdO, so the main influence is due to the chemistry.
For the $rs$ structure we compute $\varepsilon_\infty=3.16$, 5.32, 7.20 in
qualitative agreement with the reduction of the fundamental gap.
Thereby, for CdO the accuracy is reduced due to the difficulties with
the DFT-GGA band structure discussed above.

In the real parts there are several features that are similar,
independent of the polymorph.
There is only small variation of the $\varepsilon_\infty$
with the different polymorphs. In the case of ZnO we find
$\varepsilon_{\infty xx}=5.24$, $\varepsilon_{\infty zz}=5.26$ ($w$),
$\varepsilon_\infty=5.54$ ($zb$), and $\varepsilon_\infty=5.32$
($rs$). These values are larger
than the constants $\varepsilon_{\infty xx}=3.70$ and
$\varepsilon_{\infty zz}=3.75$ measured for $w$-ZnO \cite{7}. This
fact may be traced back to the underestimation of the
fundamental gap within DFT-GGA. With the values
$\varepsilon_\infty=2.94$ (measured \cite{7}) and
$\varepsilon_\infty=3.16$ (this work) the agreement is much better
for $rs$-MgO.

The lineshapes resulting from the dielectric function are
discussed in detail for ZnO (Fig.~\ref{fig:optic_zno}).
For smaller frequencies the curves
$\Re\left(\varepsilon(\omega)\right)$ exhibit maxima close to the absorption
edge. These maxima are followed by regions with the general tendency for
reduced intensity but modulated by peak structures
related to critical points in the BZ. For frequencies larger than
about 12.5~eV the real part becomes negative for all polymorphs.
For the imaginary part in Fig.~\ref{fig:optic_zno} the differences
are larger -- in agreement with the results of other calculations \cite{33}.
In the case of rocksalt there is a monotonous increase to the first main
peaks at $\hbar\omega=5.5$ and 7.0~eV which should be
shifted towards higher energies in the experimental spectra
according to the huge QP shifts discussed for the band
structure. These two peaks are due to transitions near the $L$ and $X$
points of the fcc BZ (cf. Fig.~\ref{fig:electronic_zno}). Therefore they can be
classified as $E_1$ and $E_2$ transitions \cite{59}. The
subsequent peaks at about 9.5, 12.0, and 13.2~eV should be related
to $E'_1$ (i.e., second valence-band into lowest conduction-band at $L$)
and $E'_2$ transitions. We find a different situation for the wurtzite
and zinc-blende structures. One observes a steep onset of the absorption just
for photon energies only slightly larger than the fundamental band
gap. In a range of about 4~eV it follows a more or less constant or
even concave region. Such a lineshape is clearly a consequence of
the pronounced conduction-band minimum near $\Gamma$, the nonparabolicity of the
conduction band and the light-hole valence band. It has also been
observed experimentally \cite{C}. We therefore have a rather similar
situation in the case of InN \cite{23}, the III-V compound with
constituents neighboring CdO in the periodic table of elements.\\
Within a four-band ${\bf k}\cdot{\bf p}$ Kane model
one finds for the imaginary part of the dielectric function in the
case of cubic systems \cite{23}
\begin{equation}
\label{eq:kane}
\Im\left(\varepsilon(\omega)\right)=\frac{1}{3}\left(\frac{e^2}{2a_BE_p}\right)^{\frac{1}{2}}
\sqrt{1-x}\left[\sqrt{1+x}+8\right]\theta(1-x)\bigg|_{x=E_{g}/\hbar\omega}
\end{equation}
with the Bohr radius $a_B$, the fundamental direct band gap $E_g$,
and the characteristic energy $E_p$, which is related to the square of the
momentum-operator matrix element between $s$ and $p$ valence
states. Indeed formula~\eqref{eq:kane} leads to a constant
$\Im\left(\varepsilon(\omega)\right)=3(e^2/2a_BE_p)^{\frac{1}{2}}$ for
$\hbar\omega\gg E_g$, i.e., away from the absorption edge.
Interestingly the absolute plateau values in
Figs.~\ref{fig:optic_zno}a and \ref{fig:optic_zno}b of
about 1.8 may be related to an energy $E_p$ of about $\approx37$~eV, a value which
is not too far from those in other estimations \cite{60,61}. The
same ${\bf k}\cdot{\bf p}$ model gives for the conduction-band
mass near $\Gamma$ the expression $m^*=m_0/[1+E_p/E_g]$. With an
experimental gap energy of $E_g\approx3.4$~eV it results an electron
mass of about $m^*=0.09m_0$ somewhat smaller than the experimental
value of about $0.19m_0$ \cite{61} or $0.28m_0$ \cite{7}. From the
conduction band minima plotted in Figs.~\ref{fig:electronic_zno}a and
\ref{fig:electronic_zno}b we derive band masses in DFT-GGA quality of about
$m^*=0.153m_0$ ($zb$) and $m^*=0.151m_0/0.150m_0$ $(w)$ with a negligible
anisotropy. Absorption peaks occur 
at higher photon energies at about 7.5, 9.1, and 10.6~eV for
$zb$-ZnO and 6.5, 9.9, and 10.7~eV for $w$-ZnO. These peaks may be related
to $E_1/E_2$, $E'_1$, and $E'_2$ transitions in the case of the zinc-blende structure.
For the wurtzite polymorph the interpretation is more difficult, however,
the lowest peak should mainly be caused by transitions on the $LM$ line
between the highest valence band and the lowest conduction band.

We have to mention that the
influence of many-body effects such as QP shifts
\cite{17,18,19} and excitonic effects \cite{58} has been
neglected. The QP effects lead to a noticeable blueshift
of the absorption spectra, while the excitonic effects, basically the
electron-hole pair attraction, give rise to a redshift and a more
or less strong mixing of interband transitions.
Furthermore, the neglected effects should cause a redistribution of
spectral strength from higher to lower photon energies \cite{23}.

\section{Summary and Conclusions}

Using the \emph{ab initio} density functional theory together with
a generalized gradient corrected exchange-correlation functional,
we have calculated the ground-state and
excited-state properties of several polymorphs of the group-II
oxides MgO, ZnO, and CdO. We have especially studied the rocksalt
and wurtzite structures which give rise to the equilibrium
geometries of the oxides. Zinc blende has been investigated to
study the same local tetrahedron bonding geometry but without the
resulting macroscopic anisotropy. Cesium chloride is an important
high-pressure polymorph. In addition, two hexagonal structures,
nickel arsenide and $h$-MgO, have been studied for MgO.

In agreement with experimental and other theoretical findings the
rocksalt (wurtzite) structure has been identified as the
equilibrium geometry of MgO and CdO (ZnO). The non-isostructural
series of the three compounds with a common anion has been related
to the non-monotonous variation of the cation size along the
column Mg, Zn, and Cd. We calculated binding energies which are
in good agreement with measured values.
MgO and CdO undergo a pressure-induced phase
transition from the NaCl into the CsCl structure. $w$-ZnO first
transforms into rocksalt geometry but also shows a phase
transition into the cesium chloride geometry at higher hydrostatic
pressures. Our values for the transition pressures agree
well with predictions of other \emph{ab initio} calculations and
experimental observations. We predicted the possibility of pseudomorphic
growth of MgO in biaxially strained wurtzite geometry on a $w$-ZnO
substrate.

The atomic coordination and hence the polymorph has a strong
influence on the distribution of the allowed Bloch energies. This
fact has been clearly demonstrated by the comparison of band
structures and densities of states computed for different
crystallographic structures. This fact holds in particular for the
fundamental energy gaps which however strongly suffer from the
neglect of the excitation aspect, the so-called QP
corrections. In the case of ZnO and CdO the semicore Zn3$d$ and
Cd4$d$ states also contribute to the gap shrinkage. In the
framework of DFT-GGA they are too shallow and hence give rise to
an overestimation of the $pd$ repulsion (which shifts the
uppermost $p$-like valence band towards higher energies). The
effect of the crystal structure on the dielectric function is weaker as in
the case of the band structures due to the occurring
Brillouin-zone integration. We found a stronger influence of the
cation. A clear chemical trend has been observed for the
electronic dielectric constants. Along the series MgO, ZnO, and CdO
also the averaged spectral strength increases. The crystal structure
has the most influence on the lineshape of the absorption edge. This
effect has been intensively discussed
for ZnO. For the wurtzite and zinc-blende polymorphs we observe a steep
onset in the absorption, followed by a plateau-like frequency region.

\section*{Acknowledgements}

We acknowledge financial support from the European Community in
the framework of the network of excellence NANOQUANTA (Contract
No. NMP4-CT-2004-500198) and the Deutsche Forschungsgemeinschaft
(Project No. Be1346/18-1).

\end{document}